\documentclass[structabstract]{aa}
\usepackage{graphicx}
\usepackage{txfonts}
\usepackage{natbib}
\bibpunct{(}{)}{;}{a}{}{,}
\begin{document}
   \title{Evolution of the infrared Tully--Fisher relation up to z=1.4}

   \author{M. Fern\'andez Lorenzo\inst{1}$^,$\inst{2}, J. Cepa\inst{1}$^,$\inst{2}, A. Bongiovanni\inst{1}$^,$\inst{2}, A.M. P\'erez Garc\'ia\inst{1}$^,$\inst{2}, M.A. Lara-L\'opez\inst{1}$^,$\inst{2}, M. Povi\'c\inst{1}$^,$\inst{2}, \and  M. S\'anchez-Portal\inst{3}
          }

   \institute{Instituto de Astrof\'isica de Canarias (IAC),
              C/ V\'ia L\'actea S/N, 38200 La Laguna, Spain\
   \and
     Departamento de Astrof\'isica, Universidad de La Laguna, Spain
         \and
             Herschel Science Centre, INSA/ESAC, Madrid, Spain
            }

 \date{Received.....; accepted..... }


  \abstract
   {The Tully--Fisher relation represents a connection between fundamental galaxy parameters, such as its total mass and the mass locked in stars. Therefore, the study of the evolution of this relation in the optical and infrared bands can provide valuable information about the evolution of the individual galaxies through the changes found in each band.}
   {This work aims to study the Tully--Fisher relation at high redshift in the B, V, R, I, and K$_{\rm S}$--bands by comparison with the local relations derived from a large sample of galaxies in the redshift range 0.1$<$z$<$0.3, processed in the same way, and with the same instrumental constraints that the high--redshift sample.}
   {Using the large photometric information available in the AEGIS database, we have studied the best procedure to obtain reliable k--corrections. Instrumental magnitudes are then k-- and extinction corrected and the absolute magnitudes derived, using the concordance cosmological model. The rotational velocities have been obtained from the widths of optical lines using DEEP2 spectra. At high redshift, this method is found to give better results than using the rotation curve, due to spatial resolution limitations. Finally, morphology has been determined via visual classification of the HST images. From the above information, the Tully--Fisher relations in B, V, R, I and K$_{\rm S}$--bands are derived for the local and high--redshift sample.}
   {We detect evolution in the B, V and R--bands in the sense that galaxies were brighter in the past for the same rotation velocity. The change in luminosity is more noticeable in the bluer bands. This colour evolution, unnoticed in our previous work (Fern\'andez Lorenzo et al. 2009) has been detected thanks to the more reliable k--corrections carried out in this paper, which included photometry from B to IRAC bands. The change in the (V--K$_{\rm S}$) and (R--I) colours (for a fixed velocity) could be interpreted as an ageing of the stellar populations as consequence of the star formation decrease since z=1.25. In addition, we conclude that spiral galaxies could have doubled their stellar masses in the last 8.6 Gyr.}
   {}

   \titlerunning{Evolution of the infrared Tully--Fisher relation up to z=1.4}
   \authorrunning{Fern\'andez Lorenzo et al.}
   \maketitle

%

\section{Introduction}

The relation between luminosity and maximum rotational velocity of spiral galaxies \citep{1977A&A....54..661T} is an important distance estimator, and it has been used for measuring the Hubble constant H$_0$ \citep[e.g.][]{2000ApJ..533..744}. The study of this relation at different cosmic times can probe evolution in galaxy properties such as its total mass, or the relation between dark and luminous matter \citep{1995MNRAS.273L..35Z}. Moreover, any consistent model of galaxy formation and evolution should be able to reproduce the evolution of the Tully--Fisher relation (hereafter TFR). Then, its study has important implications in determining fundamental cosmological parameters, in the study of structure formation, and in the evolution of disc galaxies \citep{2000ApJ...538..477N}.

Traditionally, the internal kinematics of nearby galaxies is measured using 21cm line widths, but the sensitivity of radio telescopes hampers its use at high redshift. Then, optical lines become important in evolutionary studies of TFR, and some authors have studied the relation between both velocity indicators. For example, \citet{1992ApJS...81..413M} compared the projected rotation velocity measured from H${\rm \alpha}$ rotation curves, with the velocity measured from integrated HI profiles at the 50\% of the maximun flux level. They obtain a difference of 10 km/s, which is attributed to the fact that HI widths measure not only the rotational velocity but the internal galaxy turbulence as well. This turbulence is important mainly in the most external region of the galaxy, where the gravity is lower. However, the optical emission does not extent so far. Then, the contribution of the turbulence to the rotational velocity is negligible when using these lines. \citet{1996vogt,1997vogt} modeled rotation curves via optical long slit observations, and they found a modest luminosity evolution in the B--band TFR ($\Delta$M$_B$(high z$-$local)$\leq$$-$0.4 at $<$z$>$=0.5). However, \citet{1997MNRAS.285..779R} and \citet{1998simard} found a stronger evolution ($\Delta$M$_B$$\approx$$-$1.5 at z=0.4), modeling the kinematics of disc galaxies in a similar way. Subsequent works have measured the evolution of the TFR, again finding different and even contradictory results. \citet{2002ApJ...564L..69Z}, and \citet{2004A&A...420...97B}, find a relation at high redshift (z$=$1) shallower than that measured in local samples, and evidence for a luminosity evolution with look--back time of ${\Delta}M_B \approx -1$ magnitudes at redshift z$=$1. In their study, contrary to previous findings, they claim that less massive galaxies were brighter in the past (at fixed rotational velocity), while the most massive would follow the local relation. However, other groups, such as \citet{2006MNRAS.366..308B}, find luminosity evolution but no slope change. Moreover, these authors find differences in luminosity evolution ranging from $-$0.2 to $-$2 magnitudes.

To cope with contributions to line widths other than that produced by rotational velocity, \citet{2006weinerI} measured kinematic line widths ($\sigma$$_{1D}$), and spatially resolved rotation and dispersion profiles. They combined line--of--sight rotation and dispersion (S$^2_{0.5}$=0.5V$^2_{rot}$+$\sigma$$^2_{2D}$), and found that this combination correlates well with the integrated line width, demostrating that $\sigma$$_{1D}$ is a robust velocity indicator, that allows constructing scaling relations with velocity for a population of diverse kinematic properties (dispersion and rotation--dominated galaxies). In another paper, \citet{2006weinerII} used $\sigma$$_{1D}$ to study the evolution in the B and J--band TFRs, finding a slope evolution in both bands, and a larger intercept evolution in the B--band ($-$1.5 mag versus $-$0.5 mag in the J--band up to z=1.2). This slope change, which implies a large evolution for the most massive galaxies, is in the opposite sense than that found by \citet{2002ApJ...564L..69Z} and \citet{2004A&A...420...97B}. Although the B--band TFR derived by \citet{2006weinerII} using resolved rotation velocities also supports an intercept evolution, the sample is too small and noisy to measure slope evolution. In addition, \citet{2007ApJ...660L..35K} demostrated that the scatter in the stellar--mass TFR is lower when using S$_{0.5}$ instead of V$_{rot}$. In \citet{2009fernandez} we analysed the optical line widths to study the evolution of the TFR in B, R and I--bands, demostrating that all optical emision lines can be used for determining galaxy kinematics. Also, we constructed our local TFRs from data derived in a similar way than the sample of high redshift galaxies, and found evidence of luminosity evolution in all three bands for the greatest redshift range of our sample, 1.1$<$z$<$1.3.

In contrast to the variety of results found in the optical bands, there seemed to be a consensus on the absence of evolution in the K--band TFR \citep{2005ApJ...628..160C,2006A&A...455..107F}. However, in a recent work, \citet{2008A&A...484..173P} do find K--band evolution in the sense that galaxies had been fainter in the past, a result opposite to what is found in the optical bands \citep[for example,][]{2006MNRAS.366..308B}. The disagreement between this work and that of \citet{2006A&A...455..107F} (both use 3D spectroscopy for deriving velocities) is due to the method for correcting the rotation velocity and, above all, the local relation used as reference. Several works have studied the local TFR in the K--band finding different results, with a slope ranging from $-$6.88 \citep{2007hammer} to $-$11.3 \citep{2001ApJ...563..694V}. Then, to obtain reliable conclusions on the evolution of the infrared TFR, is crucial to find the reason of these disagreements, fixing the local relation. Recently, \citet{2008AJ....135.1738M} presented a universal calibration of the TFR in the 2MASS J, H and K--bands. They showed that in all three bands the relation is steeper for later--type spirals, and they obtained a slope of $-$10.017 in the K--band correcting all galaxies to Sc type.

In this era of precision cosmology, the increasingly large, deep and accurate galaxy surveys within which the DEEP2 project \citep{2003SPIE.4834..161D,2007ApJ...660L...1D} provides the higher resolution galaxy spectra currently available, allow going one step forward on the study of the TFR: its colour evolution. Models of disc galaxy formation predict different colour evolution depending whether the evolution is mainly due to collapse or accretion \citep[see, for example,][]{2002A&A...389..761W}. In our previous work \citep{2009fernandez}, we studied for first time the (R$-$I) TFR evolution up to z$<$1.3, but the dispersion prevented us from distinguishing between one galaxy model from another.

In the present work, we have increased the sample to the largest redshift range 1.1$<$z$<$1.4 for all the Extended Groth Strip (EGS) with DEEP2 spectra available, with the aim to confirm or not the {\it prima facie} evidence for evolution found previously \citep{2009fernandez}. Also, we will extend our study to the V \& K$_{\rm S}$--bands, in addition to the B, R \& I--bands, in order to study the most probable galaxy formation model, since the \citet{2002A&A...389..761W} models predict larger colour differences in these bands. The local sample (0.1$<$z$<$0.3) has been extended to the whole EGS as well. Using this new local sample, we expect to set a more reliable TFR slope at z=0 to compare with our high--redshift sample.

This paper is organised as follows. In Sect. 2, a description of the data and the sample selection criteria are provided. The study of k--corrections and derivation of the absolute magnitudes and rotation velocities are described in Sect. 3. The results are presented in Sect. 4, whereas the last two sections provide the discussion of the results and conclusions. Throughout this article, the concordance cosmology with ${\Omega}_{\rm \Lambda0}=0.7$, ${\Omega}_{\rm m0}=0.3$ and $\rm H_0=70 \rm \ km \rm \ s^{-1} \rm \ Mpc^{-1}$ is assumed. All magnitudes are in the AB zero--point system.


\section{Data \& Sample Selection}

The sample consists of galaxies in the Groth Strip Survey (GSS) sky region. The baseline for spectroscopy target pre--selection were the galaxies for which DEEP2 spectra (Data Release 3, DR3) in this field were available in the redshift ranges 0.1$<$z$<$0.3 and 1.1$<$z$<$1.4. The DEEP2 project \citep{2003SPIE.4834..161D,2007ApJ...660L...1D} is a survey using the DEIMOS multi--object spectrograph \citep{2003SPIE.4841.1657F} in the Keck telescopes, to study the distant Universe. The grating used was the 1200 l/mm one, covering a spectral range of 6500--9100 {\AA} with a dispersion of 0.33 {\AA}/px, equivalent to a resolution R=$\rm \lambda$/$\rm \Delta\lambda$=4000. Despite that 1D and 2D spectra are available, in this work we only use the integrated spectra provided by DR3 \citep{1986PASP...98..609H}, due to the limited spatial resolution at high redshift. These spectra were extracted along the locus of constant lambda from the 2D spectra, using the routine {\tt do$\_$extract.pro}. Therefore, the 1D spectra provided by the DEEP2 team are corrected by effects of tilted slits.

The photometric data used here are part of AEGIS survey \citep{2007ApJ...660L...1D}. B, R, and I--band photometry were taken with the CFH12K mosaic camera \citep{2001ASPC..232..398C}, installed on the 3.6--meter Canada--France--Hawaii Telescope (CFHT). These magnitudes are included in the DEEP2 photometric catalogue  \citep[Data Release 1, DR1;][]{2004ApJ...617..765C} and the magnitude errors from sky noise and redshift are available too. The data in the V--band (F606W) were taken from HST catalogue. As IR photometry we used the K$_{\rm S}$--band, which was taken with Palomar WIRC \citep{2006bundy}. Finally, for high--z galaxies, we used data taken with the Infrared Array Camera (IRAC), on the Spitzer Space Telescope \citep{2008ApJS..177..431B} in the IRAC1 ($\rm 3.6 \ \mu m$) and IRAC2 ($\rm 4.5 \ \mu m$) bands. In Table 1, we have represented the limited magnitude and PSF (Point Spread Function) of each measurement. B, R and I--band magnitudes are already corrected for Galactic reddening based on \citet{1998ApJ...500..525S} dust maps. The V--band has been corrected following the same work. The Galactic reddening for the K$_{\rm S}$ and IRAC--bands is lower ($<$0.005 mag) than the error for these magnitudes, so they have not been considered. 

\begin{table}
\caption{Depth and average image quality of each measurement.}             
\label{table:1}      
\centering                          
\begin{tabular}{c c c}
\hline\hline
Band & Limiting Magnitude & PSF ($"$) \\\hline
B & 24.50 (8\rm $\sigma$) & 1.0 \\
V & 28.75 (5\rm $\sigma$) & 0.1 \\
R & 24.20 (8\rm $\sigma$) & 1.0 \\
I & 23.50 (8\rm $\sigma$) & 1.0 \\
K$_{\rm S}$ & 22.5 (5\rm $\sigma$) & 1.0 \\
IRAC1 & 24.00 (5\rm $\sigma$) & 1.8 \\
IRAC2 & 24.00 (5\rm $\sigma$) & 2.0 \\
\hline
\end{tabular}
\end{table}

Since in the present work we want to concentrate in the study of the TFR in V and K$_{\rm S}$--bands, we limited the sample to galaxies with these photometric bands, further selected by restricting the sample to galaxies with emission lines in their spectra, necessary for obtaining the rotation velocity. In addition, for the high--redshift sample, we only selected the galaxies with IRAC1 and IRAC2--bands photometry available, since they roughly correspond to NIR photometry at rest--frame and can provide a more reliable k--correction in the K$_{\rm S}$--band.

The inclination angle (i) must be a selection criteria as well, since no correlation between magnitude and rotation velocity can be observed for inclinations lower than 25$^{\circ}$ \citep{2009fernandez}. The inclination angle was calculated from the major to minor axes ratio as found in the HST catalogue, while the inclination errors were obtained comparing with the inclination derived using {\tt Sextractor} \citep{1996A&AS..117..393B} in the combined V+I HST images. In this way, we estimated the mean error to be ${\pm}$ 2.5$^{\circ}$ for the local sample, and ${\pm}$ 6$^{\circ}$ for the high--redshift sample. Galaxies almost edge--on will be more affected for extinction, but in our sample all galaxies have inclinations lower than 80$^{\circ}$. Therefore, the inclination of the final sample ranges between 25$^{\circ}$ and 80$^{\circ}$.

The second selection criteria was the morphology. Full details of the morphology classification are discussed in our previous work \citep{2009fernandez} to which the interested reader is referred. In order to select spiral galaxies, we performed a visual classification of every galaxy using HST images. The objects were divided into five groups: elliptical/S0 (1\%), spirals (66\%), irregulars (5\%), interacting (5\%), and unknown (23\%). To classify our visually unknown objects, {\tt GIM2D} \citep{1998ASPC..145..108S} was used. The objects with a S\'ersic index lower than 2.5 were considered as spirals. The objects visually classificated as unknown that could not be fitted with {\tt GIM2D} (5\% of "unknown" objects), were discarded. Finally, after applying all these criteria, we are left with 128 galaxies in the local sample, and 113 in the high redshift (1.1$<$z$<$1.4) sample.

\section{Data Analysis}

The luminosity and rotation velocity of disc galaxies are the parameters involved in the TFR. At high--redshift, several, not obvious, corrections are necessary to reliably obtain these parameters. Moreover, some corrections can change dramatically the results faking a real evolution. Apart from the Galactic extinction, that affects the observed magnitudes of local and high redshift galaxies in the same way, the corrections required to obtain the absolute magnitudes in each band are the k--correction and the intrinsic extinction. In this section, we will analyse the most convenient way to calculate the absolute magnitudes and kinematics of spiral galaxies. 

\subsection{Rest--Frame magnitudes}

To calculate the k--correction in the B, V, R, I and K$_{\rm S}$--bands, we need to know the spectral energy distribution (SED) of the galaxy. Since the SED is not generally known, or at least known with the required photometric accuracy, it is necessary to use an appropiate set of templates for reproducing the SED of each galaxy. Additionally, the errors in the photometric information used to fit the template and its spectral coverage can strongly affect the best--fit template and then the k--corrections. In the present work, we use the photometry included in the AEGIS catalogues. Since different instruments have been used to obtain the magnitudes in different wavelengths, the same aperture might not ensure a consistent fraction of light in each band, due to the PSF, the seeing or the pixel scale, that act spreading more or less the object. Then, we choose the photometry as close to the total magnitudes as possible, to fit the SEDs. For B, R and I--bands, \citet{2004ApJ...617..765C} measured the total R magnitudes from an aperture that systematically contains the whole galaxy flux, whereas B and I magnitudes are corrected to total ones using the colours (B--R)$_{1"}$ and (R--I)$_{1"}$ respectively. For the V--band, we choose the MAG$\_$BEST in the catalogue, which provides a nearby to total magnitude of the object. For the K$_S$--band, we used MAG$\_$AUTO because it is available for a larger number of objects than the aperture magnitudes, and we confirmed that there is no significant difference between aperture and MAG$\_$AUTO photometry due to the larger errors in this band. Finally, for the IRAC--bands, we used the MAG$\_$APER photometry in an aperture of 1.5" (only for the high redshift sample) because aperture corrections derived from average mosaic PSFs were applied to the aperture magnitudes but not to MAG$\_$AUTO, so we did not use MAG$\_$AUTO in this case. We checked that there is a good correlation between the photometry in 1.5" and 2.14" apertures, and then the whole galaxy must be inside the aperture of 1.5". 

Since in this work the photometric bands available do not match the rest--frame optical magnitudes, and the K$_{\rm S}$--band photometry is very noisy, we performed a carefull and sistematic study for establishing the most reliable procedure to do the k--correction on our data using various sets of templates and methods (see Appendix A). We found that the best result is obtained from the nonnegative linear combination of five templates based on the \citet{2003MNRAS.344.1000B} stellar evolution synthesis codes obtained by {\tt kcorrect}. However, when a noisier band is present in the data, the k--corrections calculated by this code are not suitable, since the photometric error of the observed quantity propagates to the rest--frame magnitude (Appendix A). We have found that in this case, the best method for the k--correction is obtained from photometric information that roughly match the rest--frame band for which we want to calculate the k--correction. Then, the rest--frame magnitude in the noisy band is better calculated directly from the best--fit template instead of k--correcting the observed data. This is the case for our K$_{\rm S}$--band that is roughly matched at rest--frame by the IRAC2--band. On the other hand, the k--correction in our optical bands can only be determined by interpolation, a procedure that requires available information at larger and shorter wavelengths than the band that we want to correct. Finally, we have calculated the rest--frame magnitudes used in this work convolving the rest--frame best--fit {\tt kcorrect} template with the filter response in each band.

\subsection{Intrinsic extinction}

\begin{figure*}[h!t]
\centering
\includegraphics[scale=0.40]{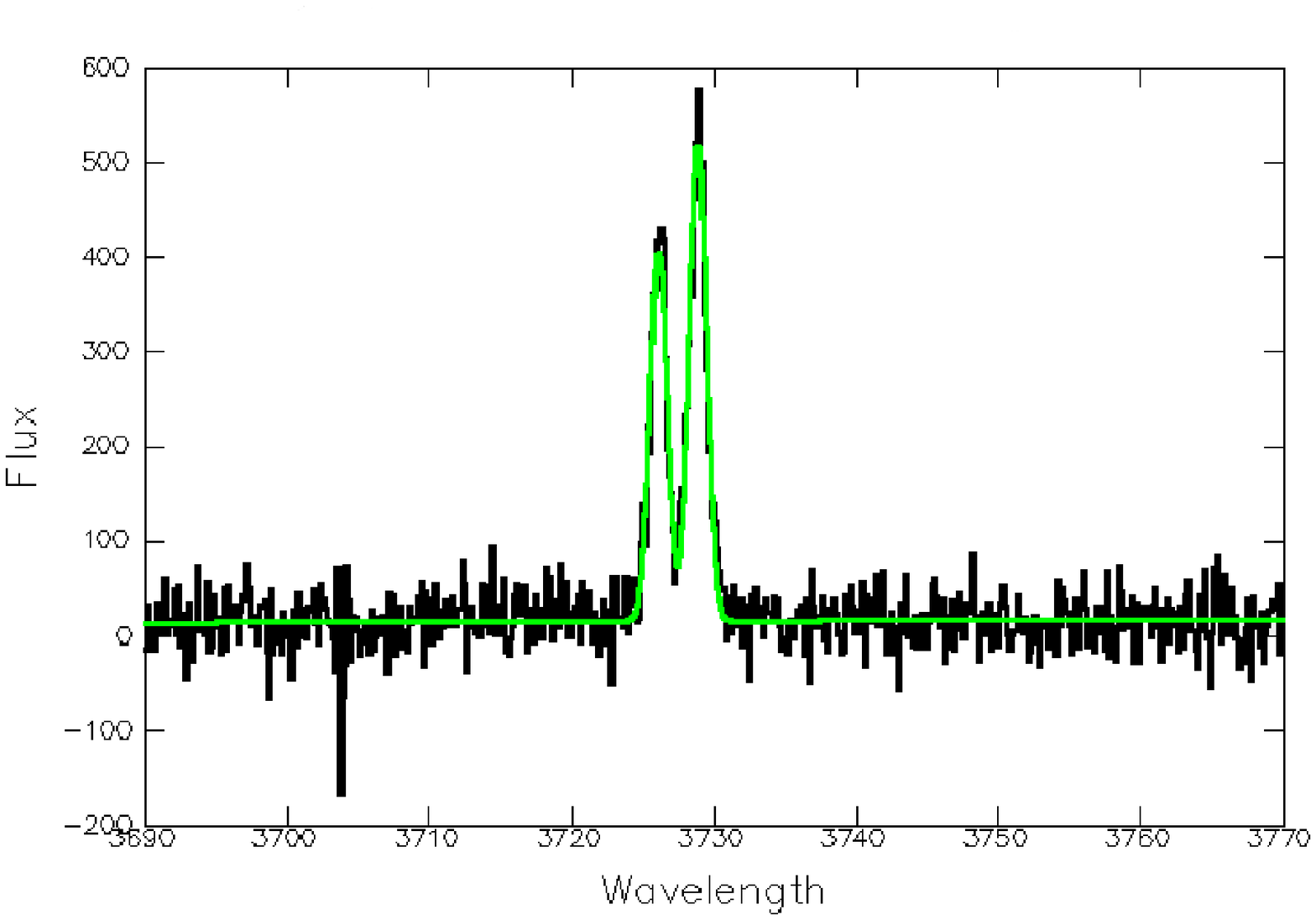}
\includegraphics[scale=0.40]{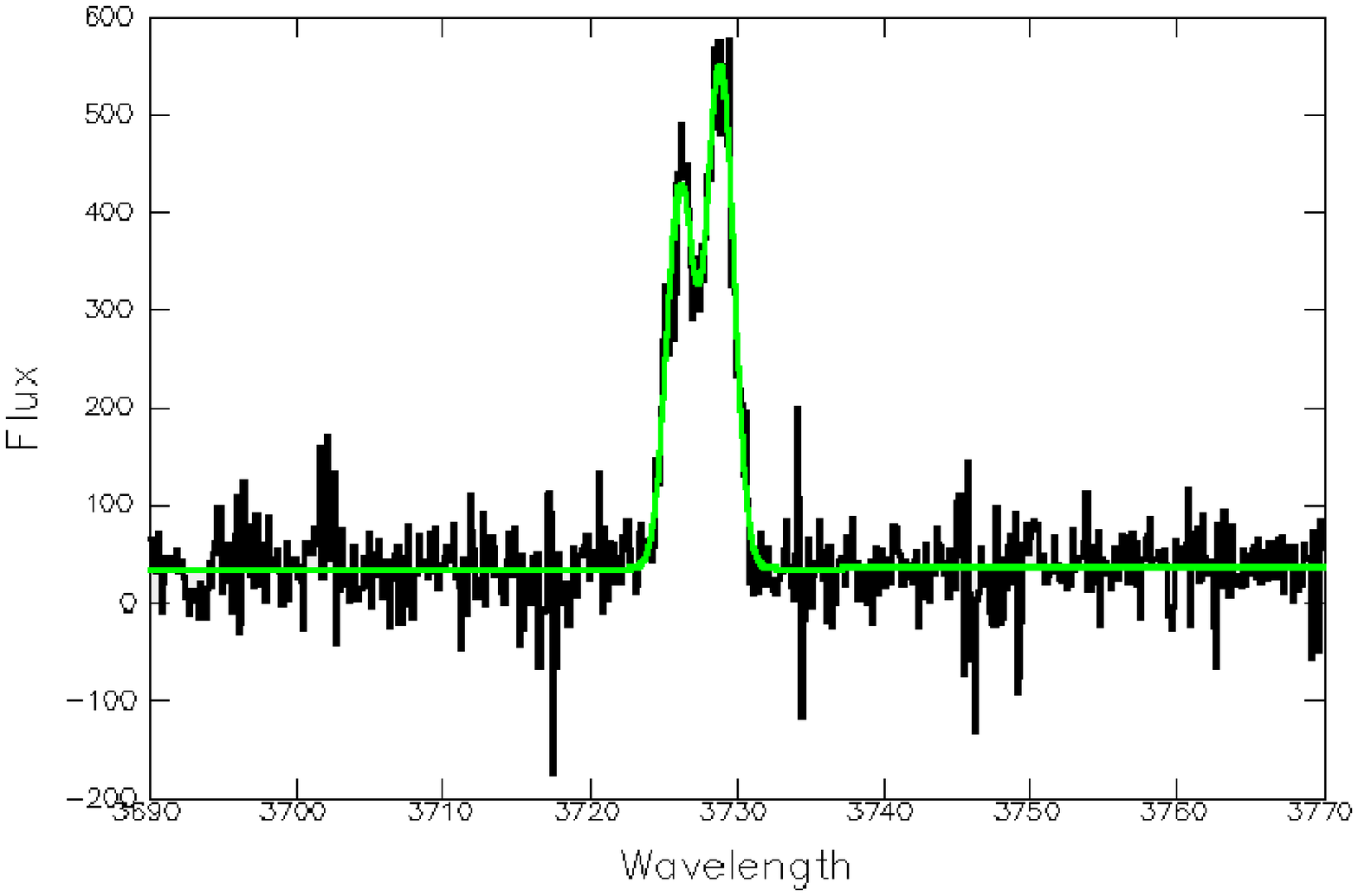}
\includegraphics[scale=0.40]{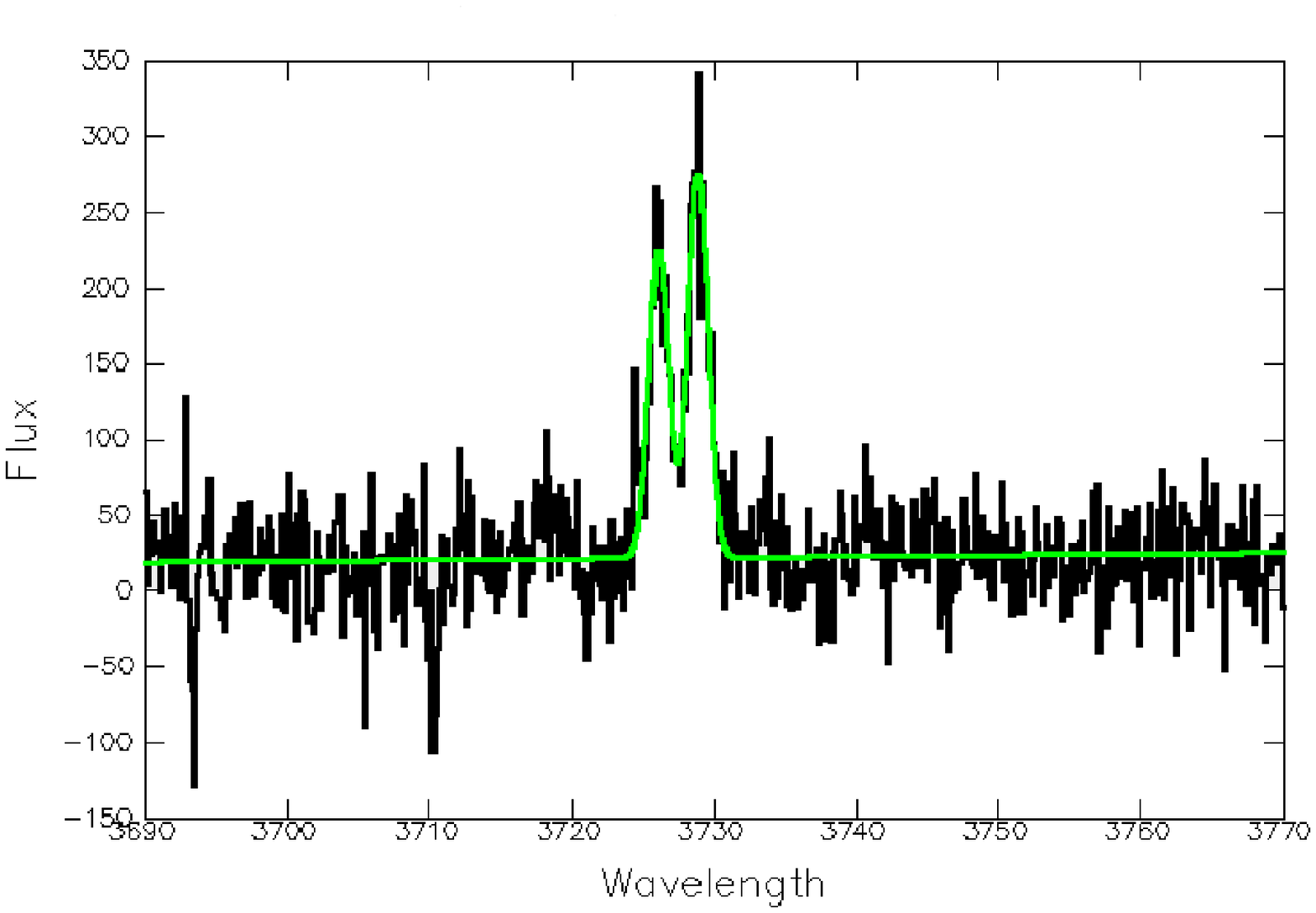}
\includegraphics[scale=0.40]{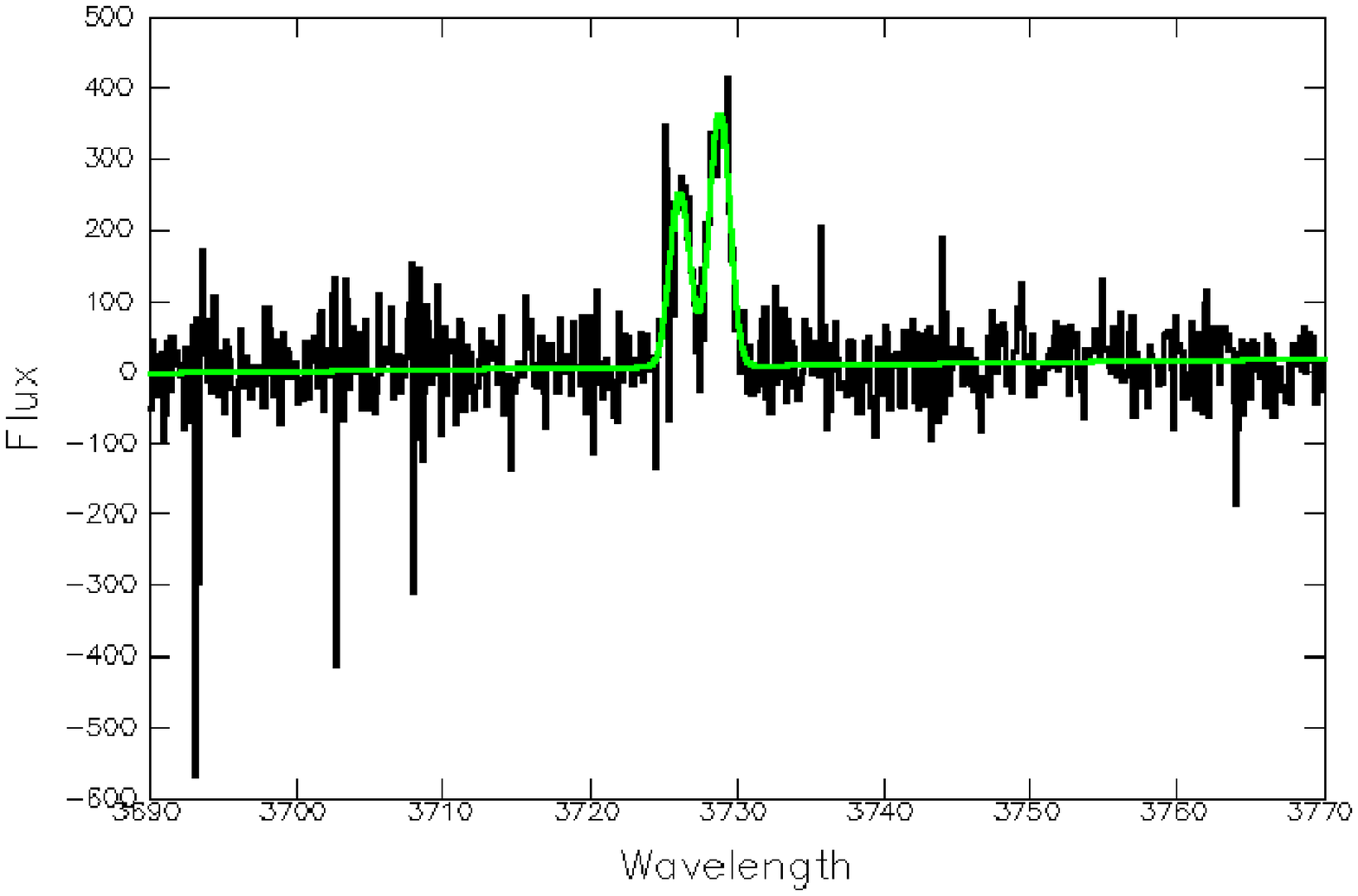}
\caption{Example spectra (not flux--calibrated) of the high--z galaxies in our sample. The x--axis represents the rest--frame wavelengths in $\AA$. The green line is the fit to the [OII]$\lambda\lambda$3727 $\AA$ double line obtained with {\tt dipso}, imposing that both lines had the same width.}
\end{figure*}

The absolute magnitudes were calculated from the luminosity distance corresponding to the measured redshifts, using the concordance cosmology. Finally, the absolute magnitudes were corrected for intrinsic extinction. This correction is basically based on inclination. In this work we have adopted the \citet{1998AJ....115.2264T} procedure, which is valid for local galaxies. Since the dust content of the galaxies may have evolved with the redshift, our extinction correction can be under or overestimated, so we will analize both posibilities in the discussion section. According to this method, the extinction $\rm A_{\lambda}$ as a function of inclination $i$ in the ${\rm {\lambda}-band}$, is defined to be:\\

\begin{equation}
A_{\lambda}^{i-0}={\gamma}_{\lambda} \ \rm log \ ({\rm a/b})
\end{equation}
with
\begin{equation}
{\gamma}_B=-0.35 \rm (15.31+M_B)
\end{equation}

\begin{equation}
{\gamma}_R=-0.24 \rm (15.91+M_R)
\end{equation}

\begin{equation}
{\gamma}_I=-0.20 \rm (16.61+M_I)
\end{equation}

\begin{equation}
{\gamma}_K=-0.045 \rm (18.01+M_K)
\end{equation}

where a/b is the galaxy major--to--minor axis ratio. These equations are valid for the magnitudes in the Vega zero--point system so our magnitudes were converted into Vega--system to do the extinction correction. For the V--band extinction, we used the Calzetti's law \citep{2000ApJ...533..682C} and we obtained $A_V^{i-0}$ = 0.8 $A_B^{i-0}$. 

\subsection{Kinematic line widths}

Most local velocities have been historically obtained from radio measurements, usually from 21cm line widths at 50 per cent of the peak intensity \citep[for example,][]{1997AJ....113...22G}. As already known \citep{1981AJ.....86.1825B}, the rotation curves are not perfectly flat at large radii. Then, since the observed HI and H${\rm \alpha}$ gas emission do not span the same radii, the velocities measured from both lines can provide different results. In addition, turbulent motions broaden the HI profile and play a role in the optical versus radio velocity width determinations \citep{1992ApJS...81..413M}. Moreover, the existence of three types of rotation curves, depending on the relation between the maximum velocity and the velocity of the flat region \citep{2001ApJ...563..694V}, can complicate the comparison. \citet{1992ApJS...81..413M} compared the rotation velocity measured from H${\rm \alpha}$ rotation curves with the velocity measured from integrated HI profiles at the 50\% level. They obtain a ratio of 2V$_c$(H${\rm \alpha}$)/W$_{50}^i$(HI) $\sim$ 0.94, which is attributed to the contribution of the turbulence to the measured HI velocity, thus overestimating the rotational velocity obtained using this line, as already mentioned. This contribution is more important in the most external region of the galaxy where the gravity is lower, while the optical emission does not spread until this region. Applying diverse corrections, \citet{1997MNRAS.285..779R} found a factor of 0.86$\pm$0.04 between 2V$_c$(H${\rm \alpha}$) and W$_{20}^i$(HI), that must be considered when comparing our results with those obtained for other local samples using 21cm.

\begin{figure*}[h!t]
\centering
\includegraphics[scale=0.44]{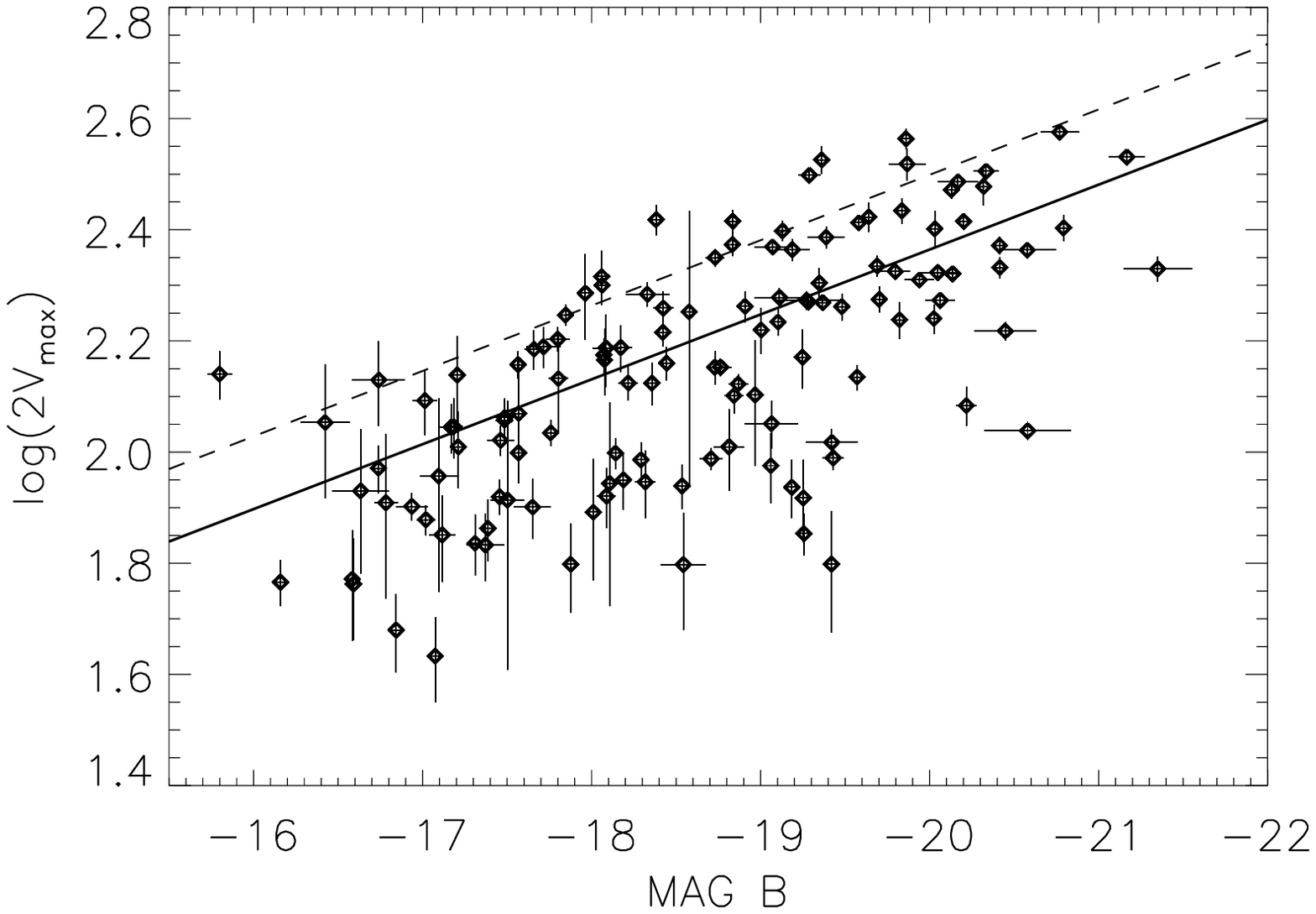}
\includegraphics[scale=0.44]{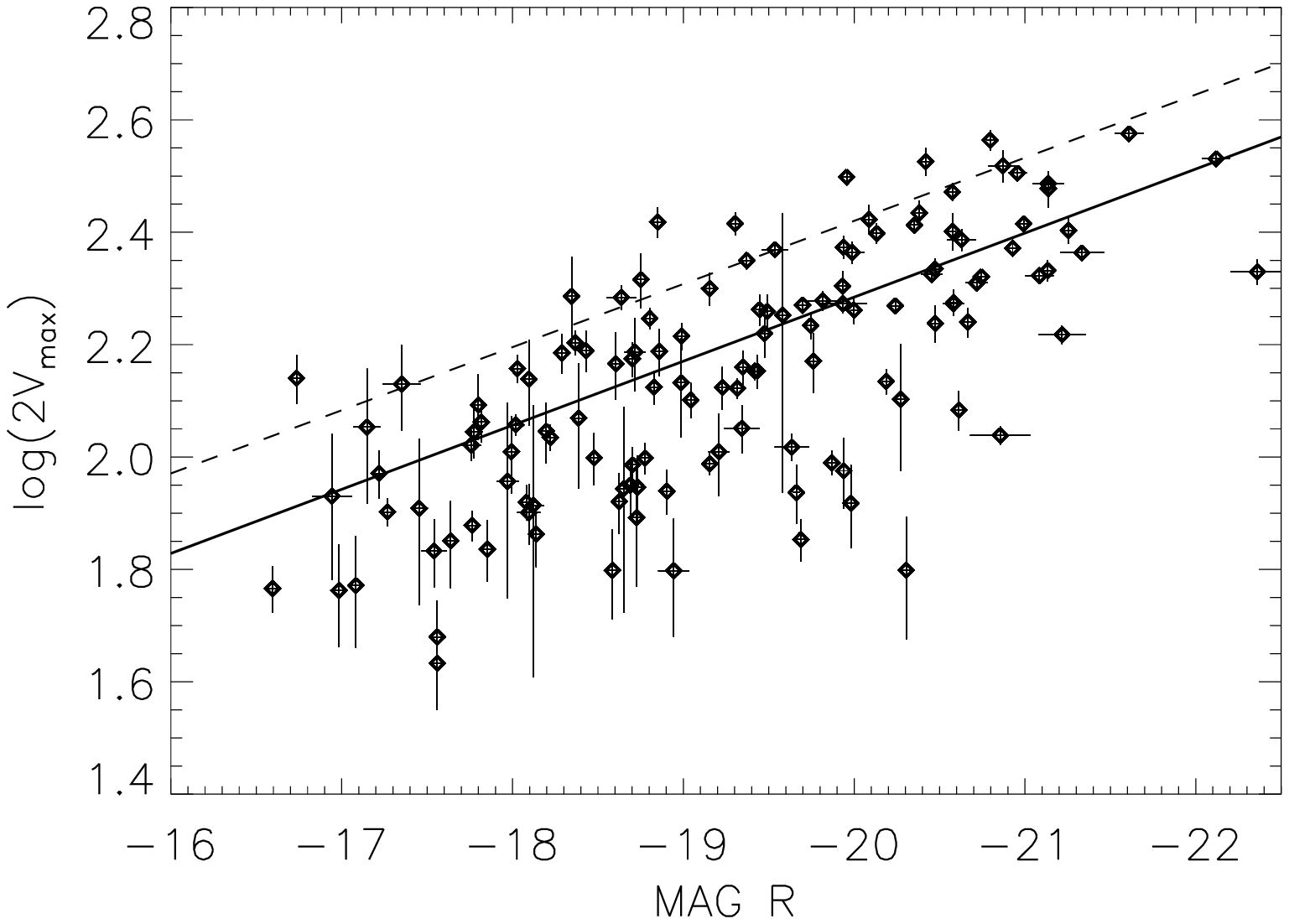}
\includegraphics[scale=0.44]{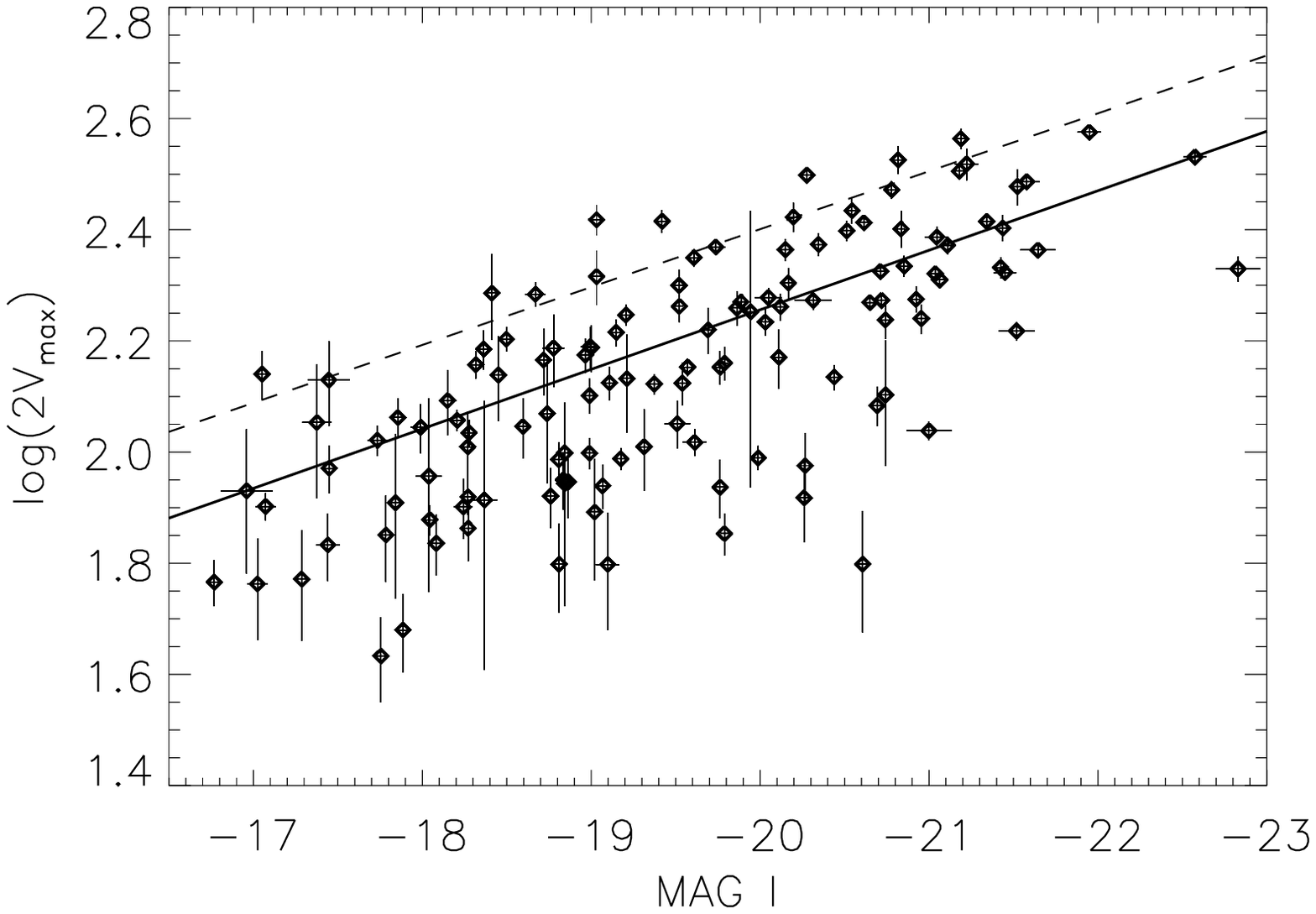}
\includegraphics[scale=0.44]{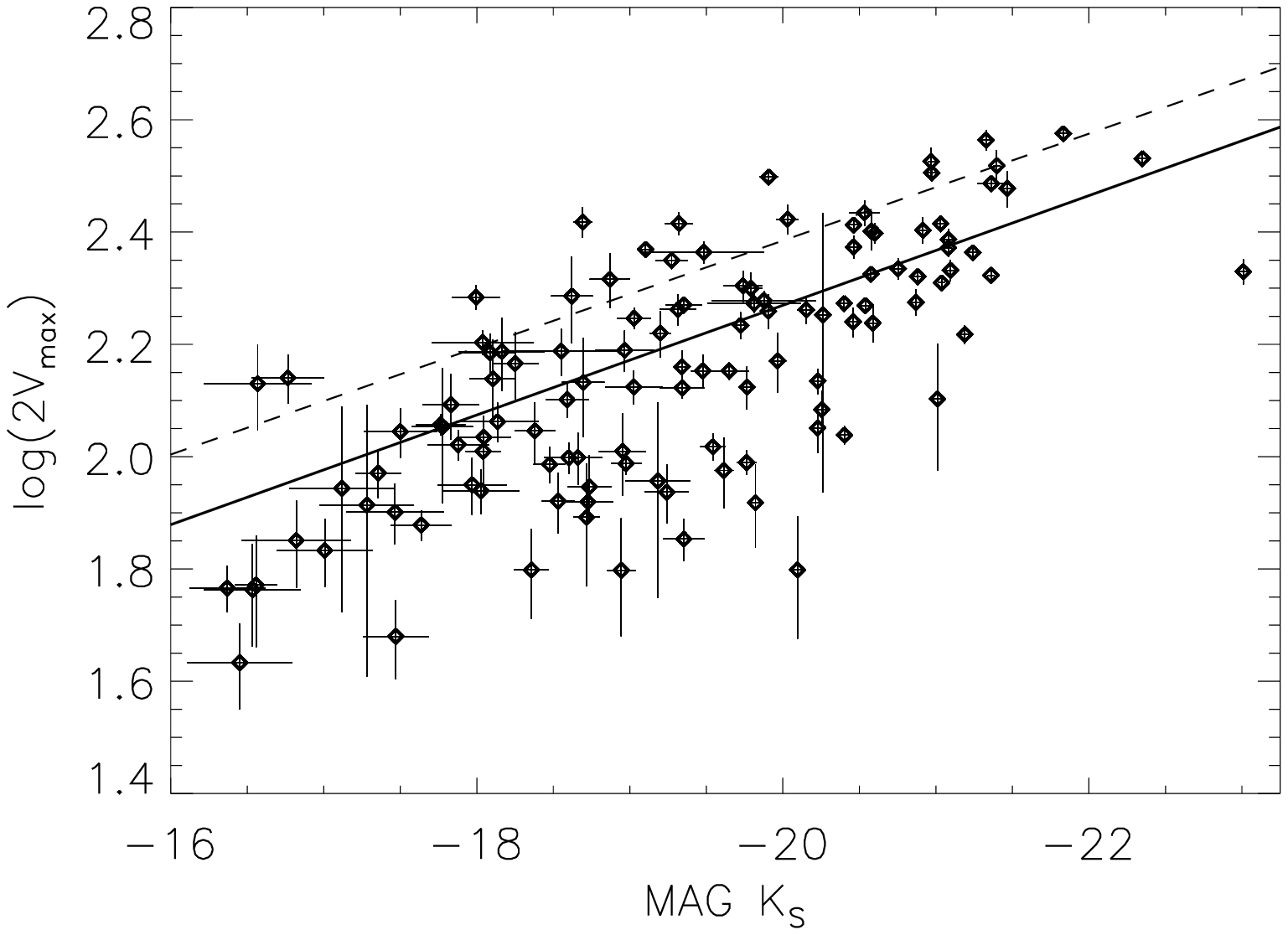}
\caption{Comparison between local TFRs obtained from DEEP2 data (0.1$<$z$<$0.3) in B, R, I and K$_{\rm S}$--bands, and \citet{2001ApJ...563..694V} local TFRs (dashed lines). The solid line represents the linear fit to the DEEP2 data. The parameters of the fit are shown in Table 2.}
\end{figure*}

In the DEEP2 survey, the long axis of every individual slitlet is oriented along the major axis of each galaxy, with an accuracy enough for deriving rotation velocities. Masks in each field are generally oriented at 2 different mask PAs, which are $-$49$^{\circ}$ and 41$^{\circ}$ in EGS. In a given mask, each slitlet can be oriented up to $\pm$30$^{\circ}$ with respect to the long axis of the slitmask. Then, from the combination of the masks and slits PAs, the maximum slit misalignments ($\delta$) would be of 11$^{\circ}$ or 19$^{\circ}$. We have checked this for a subsample of galaxies, in the ACS images, and we found good agreement between the slit and the major axis, with differences always lower than 20$^{\circ}$. This slit misalignment results in a negligible error in the velocity as already pointed out by \citet{1997AJ....113...22G} for PA offsets less then 15$^{\circ}$. Also, in a more recent work, \citet{2006weinerI} concluded that spacially resolved kinematic (2D) depends strongly on slit position angle alignment with galaxy major axis, but that integrated line width does not. For example, applying the standard cos$^{-1}$($\delta$) correction, the difference in log(V$_{\rm max}$) for a $\delta$=20$^{\circ}$ would be $\sim$0.012 for a galaxy similar to the Milky Way at z=0.2, which is lower than the error for log(V$_{\rm max}$) in our local sample. Then, we chose not to apply the slit misalignment correction.

The optical line used in this work to derive the rotation velocity is not the same for each sample of galaxies due to the limited DEEP2 spectral range. We have measured H${\rm \alpha}$ line for the local sample, and the [OII]$\lambda\lambda$3727 $\AA$ double line for the high--redshift sample. In \citet{2009fernandez}, we demonstrated that both optical emission lines can be used for determining disc--rotation velocities with the aim to compare different redshifts samples. To calculate the rotation velocity, we used:

 \begin{equation}
 2V_{max} = \frac{{\Delta}{\lambda} \ {\rm c}}{\lambda_{0}\sin{\rm (i)} \ (1+z)} \ .
 \end{equation}

where $i$ is the inclination angle, ${\lambda}_0$ the line central wavelength at z=0, and ${\Delta}{\lambda}$ the line width at 20\% of peak intensity. Equation 6 is obtained from Doppler effect, applying the corrections related to cosmology and inclination. We determined ${\Delta}{\lambda}$ with the Gaussian fitting routines in the {\tt Starlink} package {\tt dipso} \citep{1996sun50}, which calculate the FWHM (full--width half--maximum). The spectral range and resolution of the DEEP2 spectra were planned for resolving the [OII] doublet, and all our spectra fulfill this specification. In Fig 1. we have represented four cases with different S/N and line blending. To measure the [OII]$\lambda\lambda$3727 $\AA$ double line, we imposed that both lines had the same width. Then, we corrected for the instrumental width as in \citet{2009fernandez}. Where the profile is Gaussian, the line width at 20\% of the peak intensity can be compared with other velocity width measurements such as the FWHM and velocity dispersion, $\sigma$, using 

\begin{equation}
{\sigma} = \frac{FWHM}{2.35} = \frac{W_{20}}{3.62}
\end{equation}

Since our objects have been classified visually as spiral galaxies (see Section 2), we assume that all galaxies are rotating. However, we cannot prove this assumption with the existing data for the high redshift sample.

\section{Results}

\subsection{Local Relations}

In \citet{2009fernandez}, we derived the local TFRs fitting the DEEP2 data points in the redshift range 0.1$<$z$<$0.3. In this work, the local sample have been extended to all Groth field with ACS data available, with the aim to set a more reliable TFR slope at z=0 to compare with our high--redshift sample. Despite the bands that we want to study in this work are V and K$_{\rm S}$, we have derived the local TFRs in B, R and I--bands as well, in order to compare with our previous results.

\begin{table*}
\caption{Parameters of the Tully--Fisher relations obtained by fitting DEEP2 data in the redshift range 0.1$<$z$<$0.3 used as local reference. In addition, we show the values calculated by \citet{2001ApJ...563..694V} for B, R, I and K$_{\rm S}$--bands.}             
\centering                          
\begin{tabular}{c c c c c c c c}
\hline\hline
\multicolumn{1}{c}{}&\multicolumn{5}{c}{\bf This work}&\multicolumn{2}{c}{\bf \citet{2001ApJ...563..694V}}\\
Band & a & b & ${\sigma}_{total}$ & A & B & A & B \\\hline
B & $0.030\pm0.045$ & $-0.117\pm0.002$ & $0.129$ & $0.25\pm0.38$ & $-8.55\pm0.15$ & $1.24\pm0.82$ & $-8.5\pm0.4$ \\
V & $-0.026\pm0.046$ & $-0.116\pm0.002$ & $0.126$ & $-0.22\pm0.40$ & $-8.62\pm0.15$ \\
R & $0.004\pm0.042$ & $-0.114\pm0.002$ & $0.121$ & $0.04\pm0.37$ & $-8.77\pm0.15$ & $1.54\pm0.82$ & $-8.9\pm0.4$ \\
I & $0.114\pm0.039$ & $-0.107\pm0.002$ & $0.119$ & $1.07\pm0.36$ & $-9.34\pm0.17$ & $3.05\pm0.83$ & $-9.6\pm0.4$ \\
K$_{\rm S}$ & $0.316\pm0.040$ & $-0.098\pm0.002$ & $0.119$ & $3.23\pm0.41$ & $-10.20\pm0.21$ & $5.04\pm0.92$ & $-10.5\pm0.4$ \\
\hline
\end{tabular}
\end{table*}

We use an error--weighted least--squares fitting to the DEEP2 data points in the redshift range 0.1$<$z$<$0.3, to estimate the slope $a$, and the intercept $b$ of the local TFR in each band. A more detailed explanation of the fitting procedure can be found in \citet{2009fernandez}. Again, we have adopted $\log(2V_{max})$ as the dependent variable in the fit. This is the so--called inverse TFR, which is less sensitive to luminosity incompleteness bias \citep{1994ApJS...92....1W,1980AJ.....85..801S}. The results obtained for each band are given in Table 2. In addition, we show the values calculated by \citet{2001ApJ...563..694V} for the local $"$RC/FD sample (without NGC 3992)$"$ case, by fitting V$_{\rm max}$ (the same sample used for comparison in our previous work). In Fig. 2, we represented both relations in B, R, I and K$_{\rm S}$--bands. As in our previous work, the slope of the TFR in each band is very similar to that found locally by \citet{2001ApJ...563..694V}, even for the K$_{\rm S}$--band, and consistent within his errors. Moreover, our K$_{\rm S}$--band TFR slope is consistent with that derived by \citet{2008AJ....135.1738M} after correcting for incompleteness, morphology and luminosity dependence. Unlike the slope, we obtained a zero--point of the relations lower than those of \citet{2001ApJ...563..694V}. However, this difference is very similar in all bands, so it could be atributted to the true rotation velocity versus observed line width relation. \citet{1997MNRAS.285..779R} modeled simulated observations of disc velocity fields with true circular velocity V$_c$, and found $<\sigma>$/V$_c$ = 0.6, where the observed $<\sigma>$ incorporates an average over inclination. This result implies that 2V$_c$ = 0.92 W$_{20}$. In our case, a ratio of $<\sigma>$/V$_c$ $\sim$ 0.4, which provides 2V$_c$ = 1.38 W$_{20}$, would be necessary to explain the difference found in our local TFRs. \citet{1997MNRAS.285..779R} found that factors as line profile asymmetries, fiber size or inclination effects work towards reducing $<\sigma>$/V$_c$. Our spectra are not measured through fibers, but we have other effects such as misalignment of the slit, that also reduces $<\sigma>$/V$_c$. Since we are going to compare our high redshift sample with a sample of local galaxies which are measured with the same instrumentation and following the same analysis procedures, both samples are affected in a similar way by these factors, and then the differences found would be meaningful.

\subsection{High--Redshift sample}

\begin{figure*}[h!t]
\centering
\includegraphics[scale=0.44]{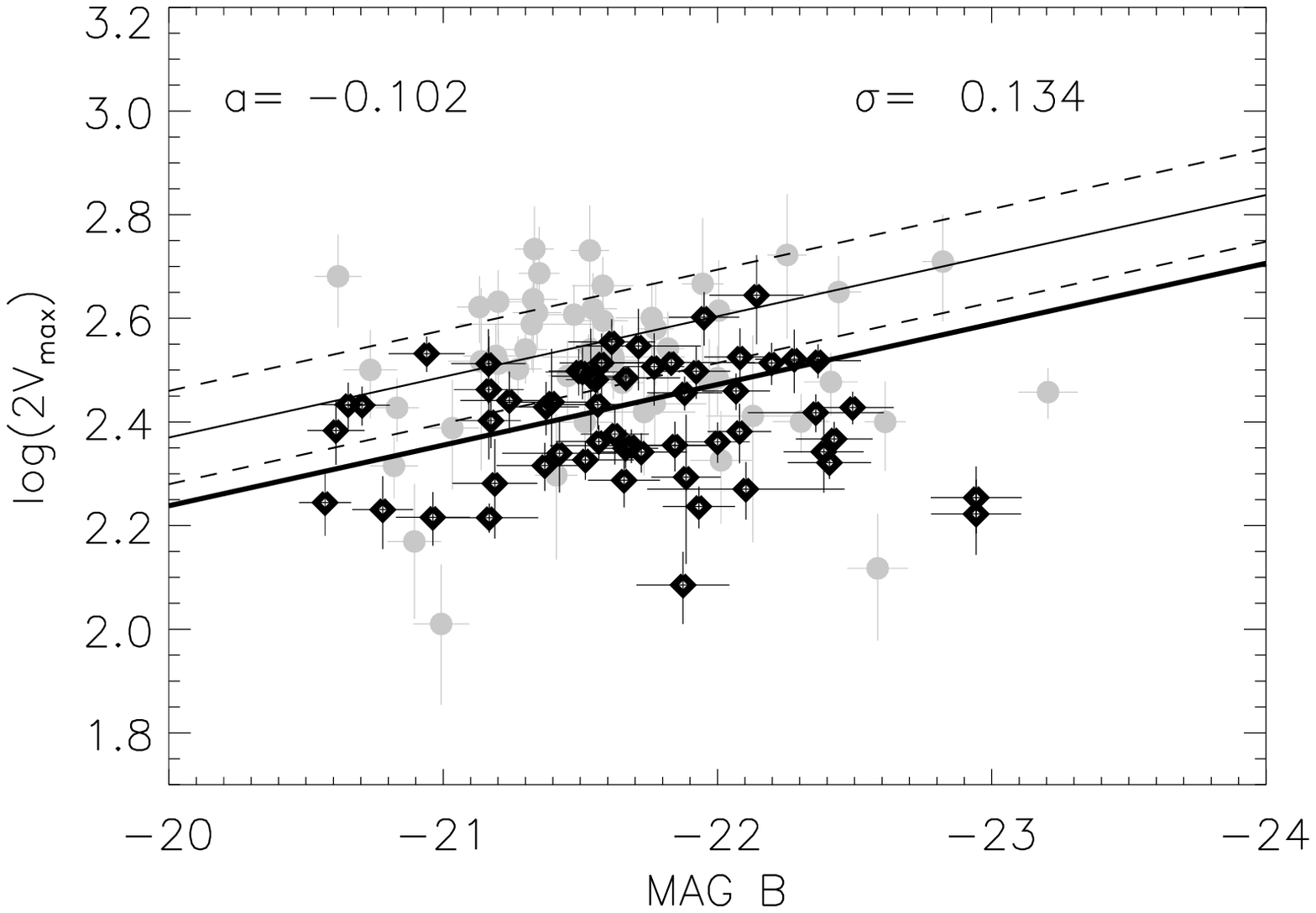}
\includegraphics[scale=0.44]{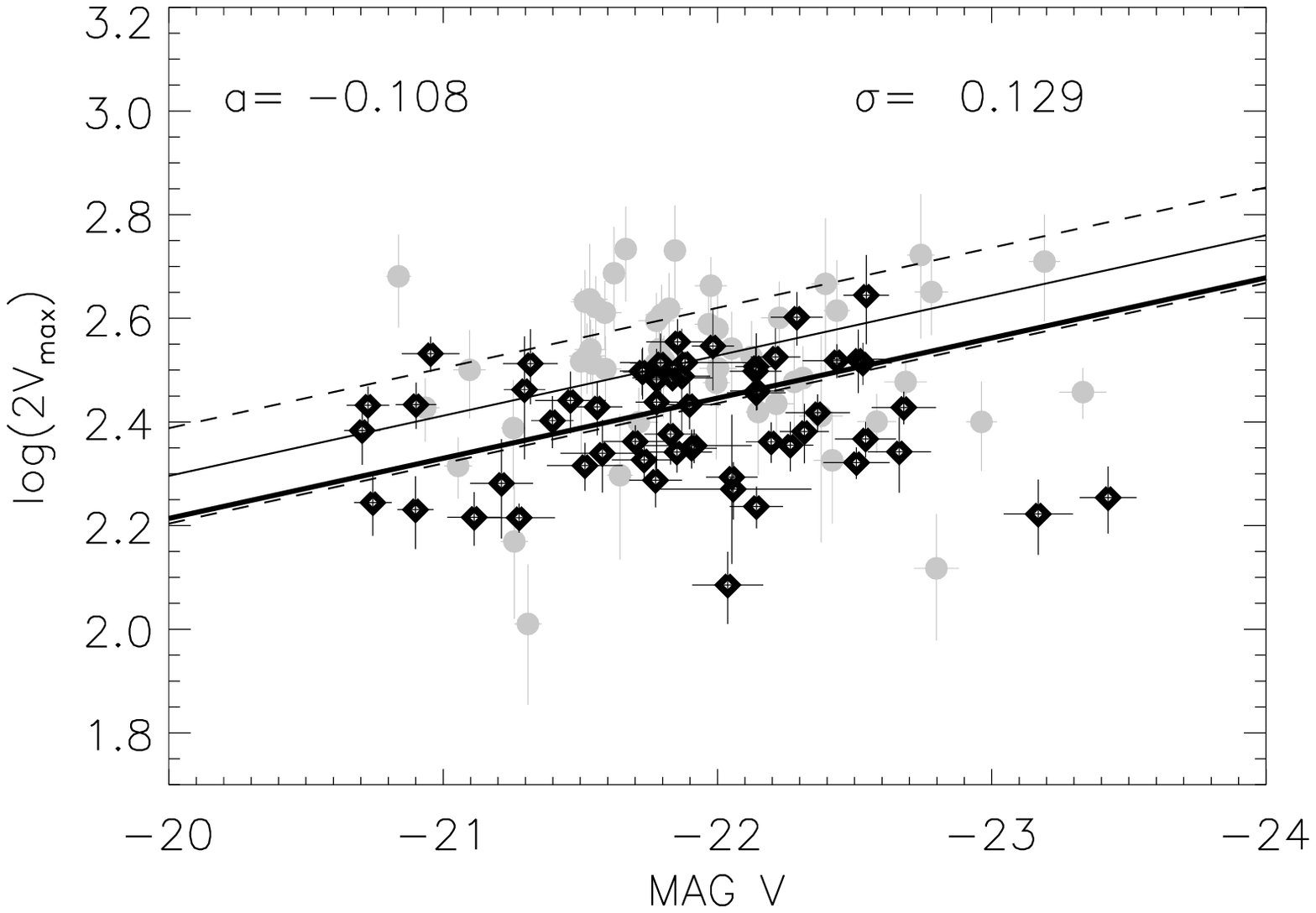}
\includegraphics[scale=0.44]{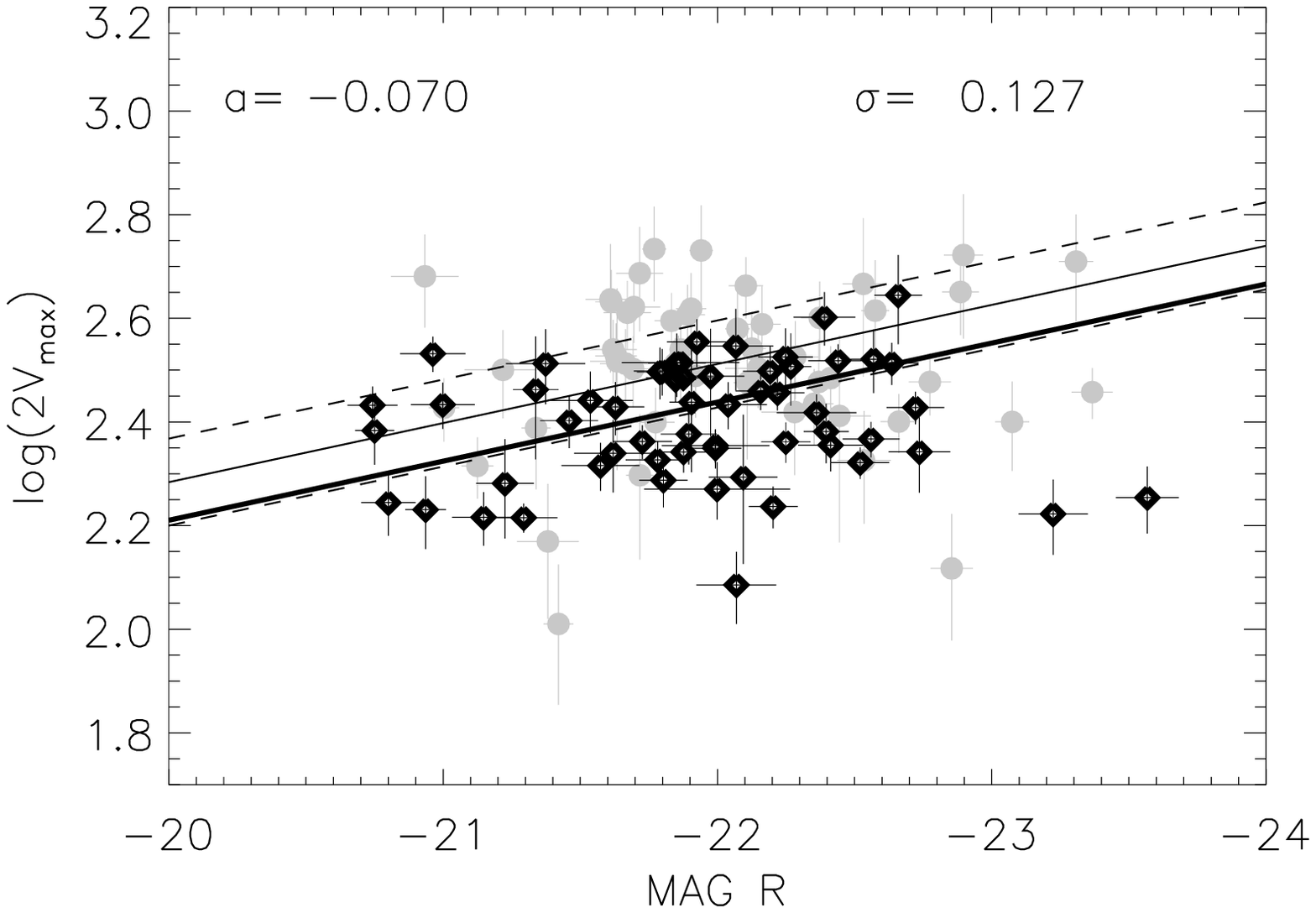}
\includegraphics[scale=0.44]{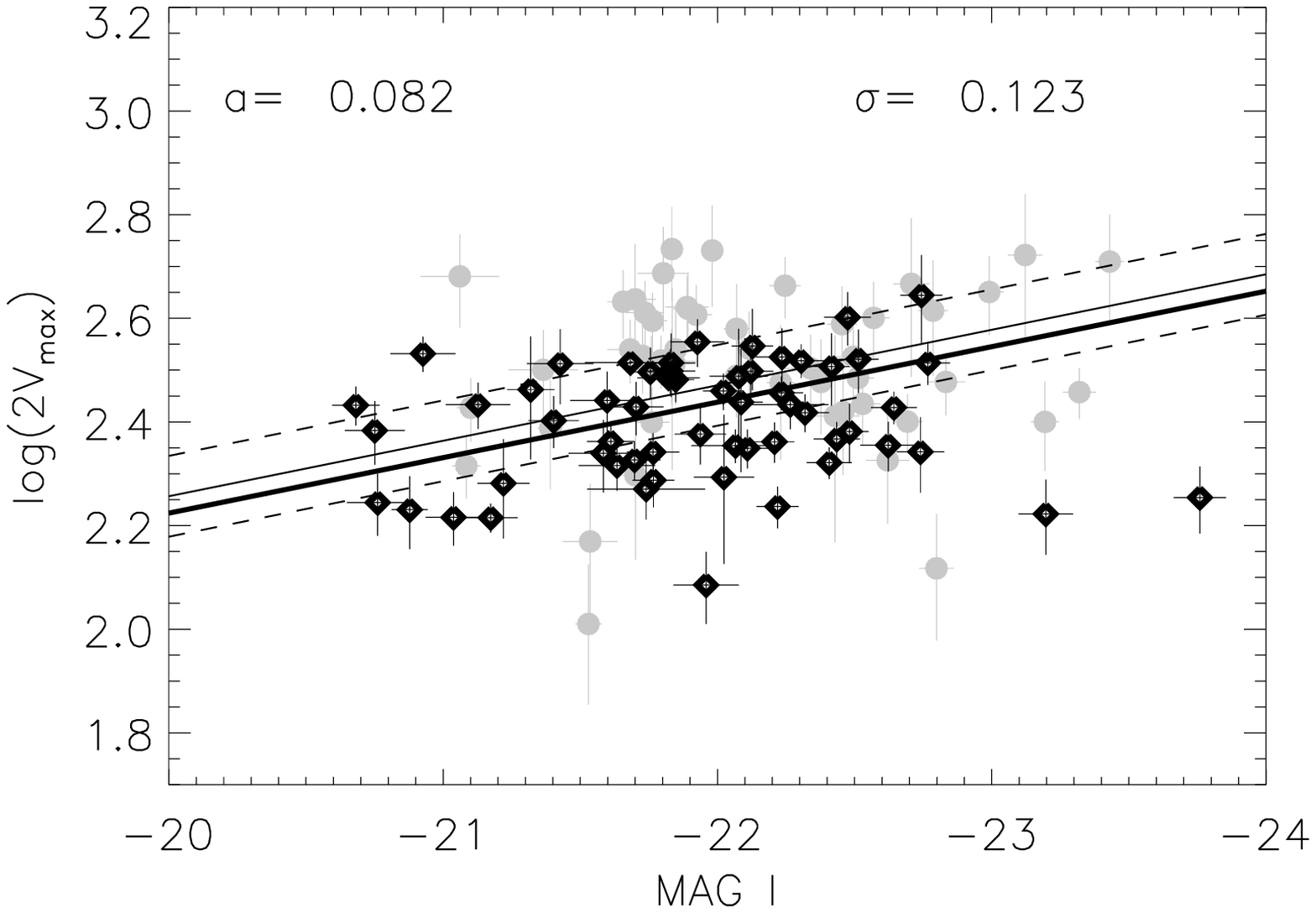}
\includegraphics[scale=0.44]{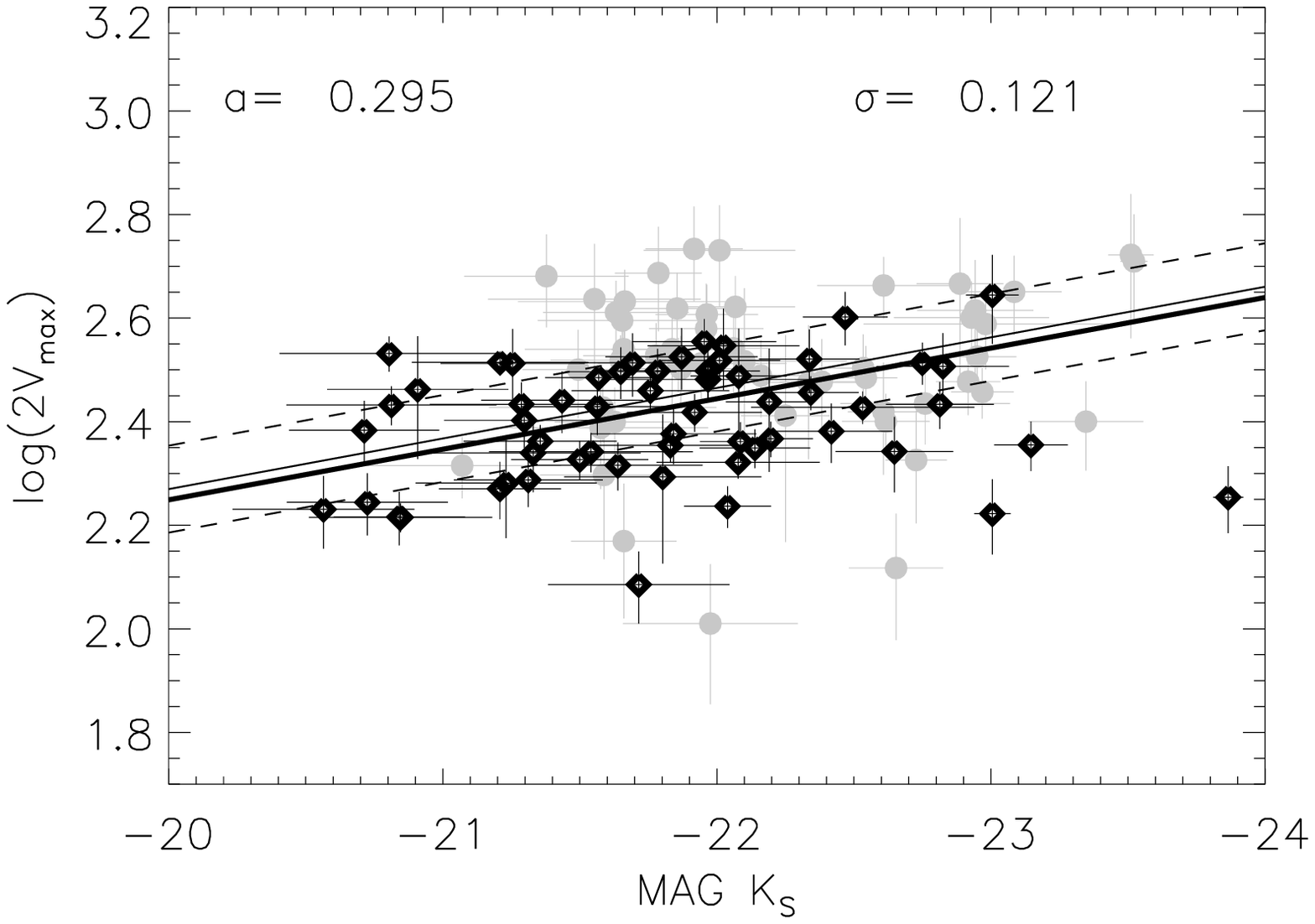}
\caption{Tully--Fisher relation in B, V, R, I and K$_{\rm S}$--bands for the redshift range 1.1$<$z$<$1.4. Black diamonds are galaxies with inclination 45$^{\circ}$$<$i$<$80$^{\circ}$, and grey points are galaxies with 25$^{\circ}$$<$i$<$45$^{\circ}$. Thin and dashed lines represent the local TFR and its 2$\sigma$ uncertainty in the offset. The thick line is the weighted least--squares fit to all points. The fit was performed by fixing the slope to the local one in each case.}
\end{figure*}

In Fig. 3 we have represented B, V, R, I and K$_{\rm S}$--band TFRs, for the redshift range 1.1$<$z$<$1.4, to test the evidence of evolution found in our previous work but with a larger sample of galaxies. In order to examine the intercept evolution we have set the slope to the local one of the TFR in each band. The large scatter of the high--z relation is noticeably reduced when we limit the sample to galaxies with inclination $i>45^{\circ}$ (see Fig. 3), which is the lower value used in some works \citep[for example,][]{2006MNRAS.366..144N}. If we fit separately objects with $i<45^{\circ}$, we found lower evolution than for all objects together. For galaxies with $i>45^{\circ}$ the opposite happens, i.e. we found larger evolution in the zero--point. For galaxies with low inclination, an error in the inclination angle yields larger velocity errors than for more inclined galaxies. Nevertheless, the error in the extinction will be more noticeable for galaxies with high inclination. If the error in the extinction were the reason of the difference between both inclination ranges, then for the K$_{\rm S}$--band there would be no change between the evolution found for galaxies with low and high inclination, since the extinction is negligible in this band. However, when both inclination ranges were fitted independently for the K$_{\rm S}$--band, we found the same difference that for the other bands. Therefore, a velocity error due to an inclination error seems to be the reason of the disagreement between both inclination ranges. Although in terms of the velocity, the results are more reliable for galaxies with $i>45^{\circ}$, the evolution in the zero--point of the TFRs do not change significantly if the whole sample is used. Therefore, we fitted all galaxies together to increase, into the bargain, the statistical significance of the result. The scatter in the TFR and its evolution are important to constraint models of galaxy formation. \citet{2008AJ....135.1738M} observed that the scatter in the local K--band TFR increases with decreasing rotation velocity. In our local sample, the scatter is 1.21 mag in the K--band, but it is reduced to 0.9 mag when we select the objects with log(2Vmax)$>$2.2. For log(2Vmax)=2.2, the scatter in the K--band TFR obtained from \citet{2008AJ....135.1738M} is $\sim$0.9 mag. However, the scatter in our high--redshift sample (1.23 mag) is larger than the scatter of $\sim$0.5 mag obtained in \citet{2008AJ....135.1738M} for log(2Vmax)=2.5, that is the average velocity of our high redshift sample. Then, although the scatter does not evolve significantly with respect to our local relations (consistent with the result found by \citet{2005ApJ...628..160C} for their samples at z$<$0.7 and z$>$0.7), it is two to three times higher when comparing with the local relations of \citet{2001ApJ...563..694V}, \citet{2008AJ....135.1738M}, or \citet{2007hammer}. This is probably due to measurement errors larger than the errors for z=0 galaxies.

\begin{table}
\caption{Magnitude evolution found in each band assuming a constant slope.}             
\label{table:1}      
\centering                          
\begin{tabular}{c c c}
\hline\hline
Band & $\Delta$M & ${\sigma}_{{\Delta}M}$ \\\hline
B & -1.125 & 0.535 \\
V & -0.708 & 0.547 \\
R & -0.646 & 0.531 \\
I & -0.298 & 0.545 \\
K$_{\rm S}$ & -0.213 & 0.610 \\
\hline
\end{tabular}
\end{table}

As we have assumed that the slope does not change with the redshift, we can calculate the evolution of the magnitude for each band, via the difference between the high--redshift and local TFRs for a fixed velocity. The results are shown in Table 3. For the B--band, we found a change in magnitude of $\rm \Delta$M$_{\rm B}$=$-$1.13 at z=1.25, which is larger than 2$\sigma$, and consistent with our previous work \citep{2009fernandez}, and that found by other authors \citep[e.g.,][]{2004A&A...420...97B,2006MNRAS.366..308B}. However, for R and I--band, we found weaker evidences for evolution, that likely can be attributed to the more reliable k--corrections carried out in this work. The most interesting result is a lower difference in the intercept as the band is redder, being the change in the zero--point for I and K$_{\rm S}$--bands less than 1$\sigma$. The difference in magnitude found for the K$_{\rm S}$--band is compatible with no evolution, and it is in agreement with the results found by \citet{2005ApJ...628..160C} and \citet{2006A&A...455..107F}. Nevertheless, \citet{2008A&A...484..173P} found evolution in the K$_{\rm S}$--band TFR, in the sense that galaxies have been fainter in the past. They attributed the disagreement with the result of \citet{2006A&A...455..107F} to both the local relation slope used as reference and the more accurate measurement of the rotation velocities of \citet{2008A&A...484..173P}. However, if the \citet{2008A&A...484..173P} data are compared with the \citet{2001ApJ...563..694V} local relation \citep[the same used by][]{2006A&A...455..107F}, no evolution is found. Therefore, the local TFR slope is playing a fundamental role in the evolution of the K$_{\rm S}$--band TFR, and it will be studied separately in the next section.

\section{Discussion}

Several authors have studied the infrared TFR both in the K and in the K$_{\rm S}$--bands. To compare with these works, we need to know the relation between both bands. From \citet{2002AJ....123.1603G}, a Bessell K--band is related with the 2MASS K$_{\rm S}$--band as $\rm K=K_S-(-0.044\pm0.003)-(0.000\pm0.005)(J-K)$. Assuming that 2MASS and Palomar K$_{\rm S}$--bands are equivalent, the difference between K and K$_{\rm S}$--bands is within the errors in our K$_{\rm S}$--band absolute magnitudes, and therefore we can compare our results with those obtained in other works.

\subsection{The slope in the local K$_{\rm S}$--band TFR}

\begin{figure*}[h!t]
\centering
\includegraphics[scale=0.44]{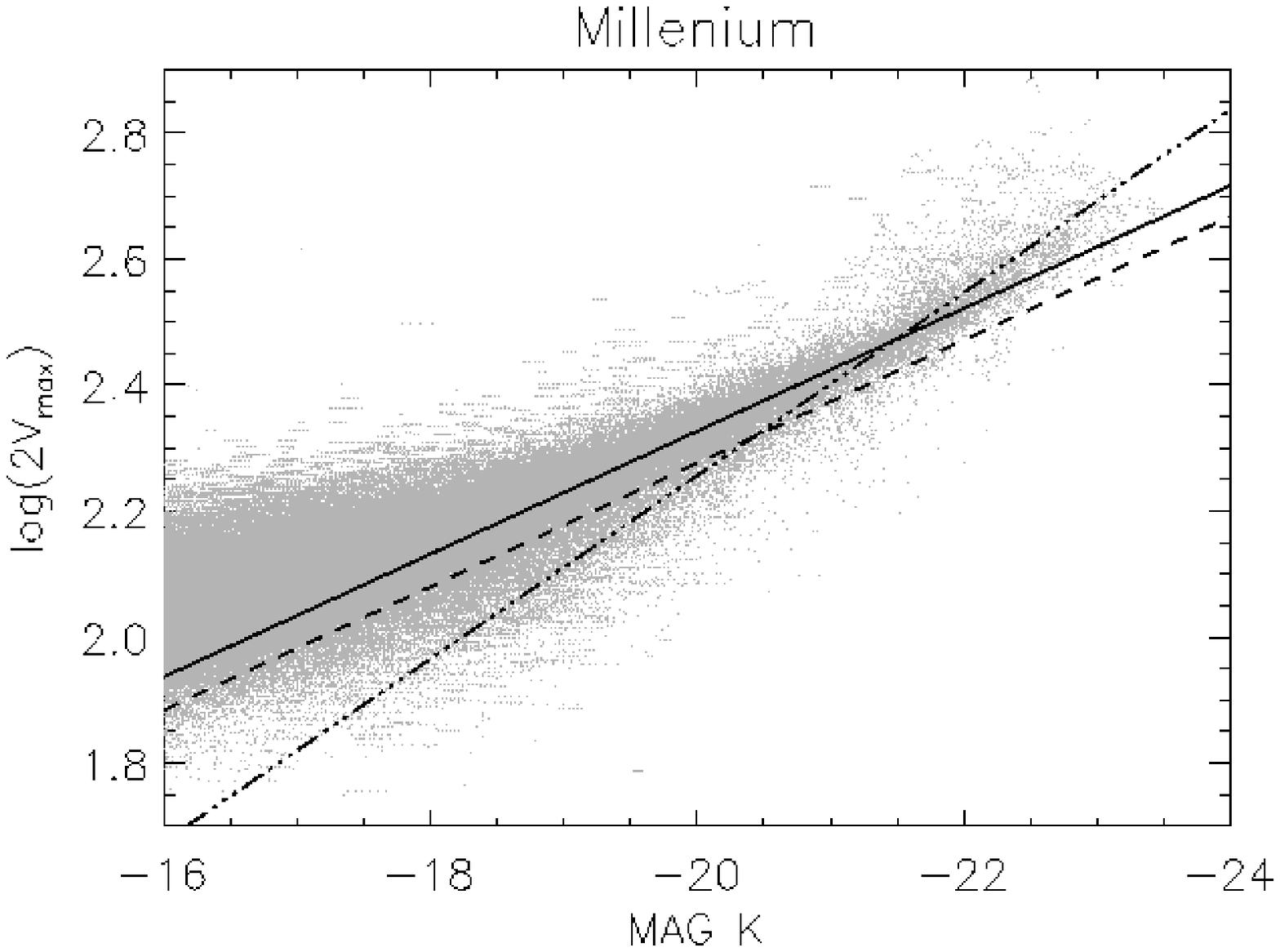}
\includegraphics[scale=0.44]{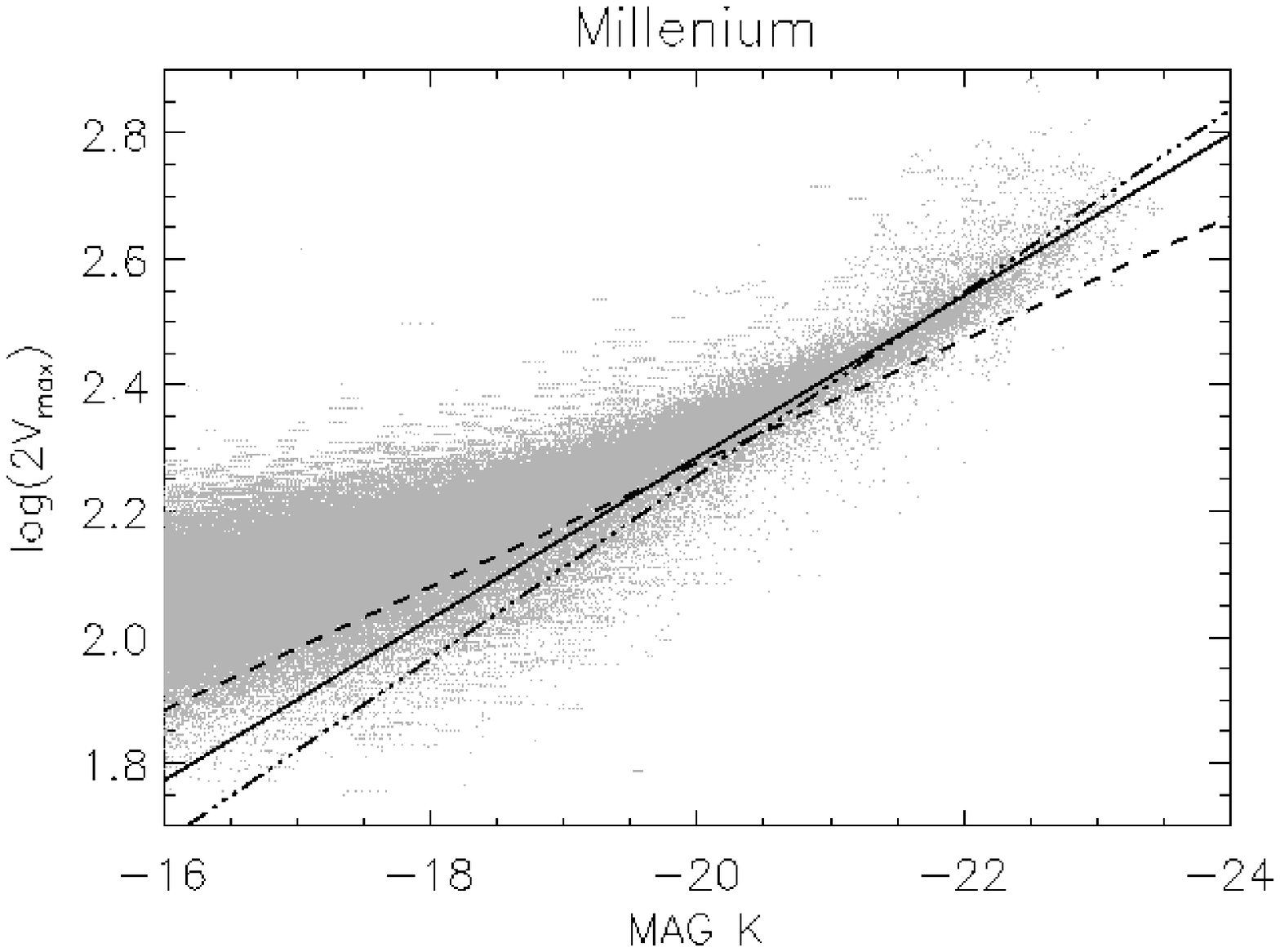}
\caption{Local Tully--Fisher relation in the K$_{\rm S}$--band for the Millenium mock galaxies. The grey points are the objects with $\rm M_{bulge}$ $<$ 0.15 $\rm M_{star}$ and $\rm M_K$$<$ $-$16. The dashed line is the relation obtained with our local sample and the three dot--dash line is the local relation obtained by \citet{2007hammer} for a sample of SDSS galaxies with log(V$_{flat})\geq$2.2. The solid line is the fit to the data for M$_K$$<-$18 (left) and M$_K$$<-$21 (right).}
\end{figure*}

In order to study the discrepancies found by the authors in the slope of the local K$_{\rm S}$--band TFR, we use the Millenium Simulations database \citep{2005Natur.435..629S} carried out by the Virgo Consortium. From this simulation, the catalogue that we used, contained in the {\tt millimil} database, is the table DeLucia2006a, from which we took the data corresponding to maximum rotational velocity (v$_{\rm max}$); Johnson B, V, R, I, and K absolute rest--frame magnitudes (dust extinction included); redshift; mass of bulge (M$_{\rm bulge}$), and the total mass in stars (M$_{\rm star}$) \citep{2007MNRAS.375....2D,2006MNRAS.365...11C}. Using the upper value of bulge--to--total (B/T) mass ratio for late--type galaxies from \citet{2006laurikainen}, we select the objects with $\rm M_{bulge}$ $<$ 0.15 $\rm M_{star}$ as representative of spiral galaxies. In Fig. 4 we represent the data for $\rm M_K$$<$ $-$16. K--band magnitudes have been converted into the AB system using M$_K$(AB)=M$_K$(Vega)+1.9 \citep{2006hewett}. The slope of the local TFR obtained using simulated galaxies seems to change with the magnitude range considered. For instance, when we fit galaxies with M$_K$$<$ $-$18, we obtain a slope very close to our sample of local galaxies. However, for M$_K$$<$ $-$21, the slope is closer to the K--band TFR found by \citet{2007hammer} restricting the sample to SDSS galaxies with log(V$_{flat})\geq$2.2 \citep[the local TFR used by][]{2008A&A...484..173P}. For the other bands, we found the same tendency for the slope as that found in K--band. 

If we assume that the slopes of the TFRs change with the magnitude range considered, then our local relations cannot be compared with our high redshift sample, which consists of galaxies brighter than M$_K$ = $-$20.5. In order to compare our high redshift data with a local relation derived for the same magnitude range, we make use of the Millenium data. With this aim, we calculate the difference in the zero--point between our local sample and Millenium data by setting the same slope in the same magnitude range as in our local one. As it happens in the comparison with \citet{2001ApJ...563..694V} local relations, we obtained a similar difference in the intercept of the relation for all bands, that can be attributed to the ratio between rotation velocity and line width. Nevertheless, in this case, the difference in the zero--point is lower, being the zero--point of the \citet{2001ApJ...563..694V} relations greater than the Millenium local zero--points. Then, we calculated the Millenium local TFR relations for the brighest galaxies and we apply the shift in the zero--point necessary to compare with our high redshift sample. Finally, we fit the data of the high redshift sample by setting the slope to the new local one determined using Millenium data. We found a larger evolution in luminosity for all bands, but for I and K$_{\rm S}$--bands, the difference in the zero--point of the high redshift relation with respect to the Millenium local one remains inside 2$\sigma$ (where sigma is the uncertainty in the local offset), being $\sim$1$\sigma$ for the K$_{\rm S}$--band. Therefore, we found a lower luminosity evolution as passing from blue to redder bands, consistent with the results obtained in the present work. Then the no evolution of the TFR observed in the K$_{\rm S}$--band is reinforced.

\subsection{The colour evolution of disc galaxies}

The colours of galaxies provide information on their stellar content and, through evolutionary models, on the history of star formation. Then, the colour evolution versus rotational velocity, allows us to study the change in the stellar content of a galaxy with respect to its total mass. In Fig. 5 we represented the colours (V--K$_{\rm S}$) and (R--I) versus rotational velocity for local and high--redshift galaxies. For (V--K$_{\rm S}$) we have a larger dispersion, and the evolution of $\sim$0.35 mag found by setting the slope to the theoretical local (V--K$_{\rm S}$) relation, is within the dispersion (less than 1${\sigma}$). However, there is a clear change of $\sim$0.33 mag with the redshift, in the (R--I) colour at fixed velocity ($>$2${\sigma}$). Then, the galaxies are redder today, which could be due to an ageing of the stellar population as consequence of a star formation decrease in the last 8.6 Gyr. As in the case of the evolution in R and I--band TFRs, we found a different result than that found in \citet{2009fernandez}, also due to the more reliable k--corrections carried out in this work, where we have used additional photometric information.

\begin{figure}[h!t]
\centering
\includegraphics[scale=0.35]{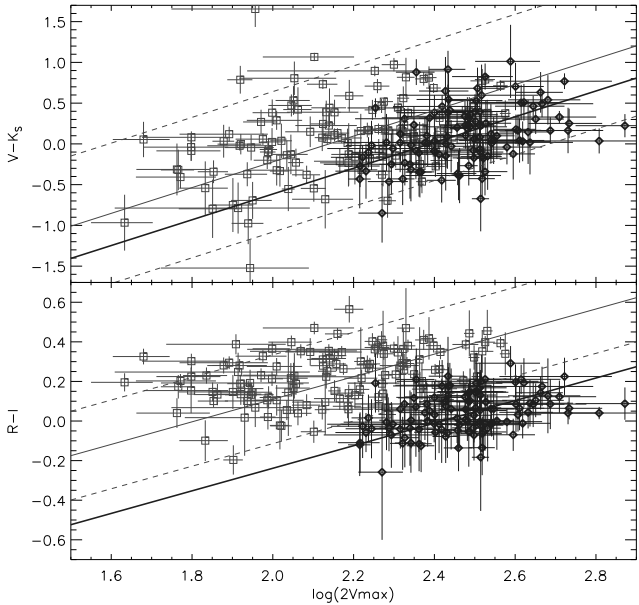}
\caption{Colour (V--K$_{\rm S}$) vs. rotation velocity (up) and colour (R--I) vs. rotation velocity (down). Black diamonds represent the galaxies in the redshift range 1.1$<$z$<$1.4, while open squares are the galaxies in the redshift range 0.1$<$z$<$0.3. Difference between the local (V--K$_{\rm S}$) TFRs (up) and (R--I) TFRs (down) (thin lines) and its 2${\sigma}$ uncertainty (dashed lines) are shown. Thick lines are the fit to the high redshift data setting the slope to the local relation in each case.}
\end{figure}

The colour evolution of simulated disc galaxies with redshift has been investigated in some works. In \citet{2009fernandez} we compared the results with \citet{2002A&A...389..761W}, which studied the colour evolution of disc galaxies in two models of galaxy formation: the accretion and the collapse model. Using these models, the change in colours observed here would be better reproduced by the accretion model, but the predicted evolution in the (V--K) colour would be double than that found here, and the change in magnitude is opposite to that predicted by the simulations. Moreover, in a subsequent paper, \citet{2007A&A...465..417W} investigated the effect of the Inicial Mass Function (IMF) on the colour evolution of disc galaxies using two different IMFs: Salpeter and Kroupa. The evolution in the (g--K) colour derived from their work (for the Kroupa IMF without absortion) would again be larger than the result found here, assuming (g--K)$\approx$(V--K$_{\rm S}$), mainly due to the change in the K--band of $\sim$1 mag predicted by the simulations. But this change in K would lead, according to their evolution of the M$_{\ast}$/L$_{\rm K}$, to a growth of a factor $\sim$6 in the stellar mass since z=1.25. However, from the M$_{\ast}$/L$_{\rm r}$ and M$_{\ast}$/L$_{\rm i}$ given in the same work, this change in the stellar mass would represent an evolution of $\rm \Delta M_r$=0.36 and $\rm \Delta M_i$=0.26 from z=1.25 to z=0, which would imply that galaxies were fainter in the past. This result is opposite to that found here, and that found by \citet{2006A&A...450...25V} for the R--band. Then, our results cannot be reproduced by the simulations of \citet{2002A&A...389..761W,2007A&A...465..417W}

\subsection{Comparison with other works}

As we discuss in \citet{2009fernandez}, the evolution found in the optical bands is more likely due to a change in luminosity than to a change in velocity. \citet{2005ApJ...628..160C} and \citet{2006A&A...455..107F} obtained the same result than in this work, i.e. no evolution in the K--band TFR, whereas \citet{2008A&A...484..173P} found a change of the K--band TFR zero--point, which they attributed to an average brightening of galaxies since z$\sim$0.6 by 0.66$\pm$0.14 mag. They explain this brightening as a growth in stellar--mass by a factor $\sim$2.5 since z=0.6, as evaluated from the evolution in log(M$_{stellar}$/L$_K$) found by \citet{2004ApJ...608..742D}. They claim that the result is consistent with the gaseous O/H phase abundance of z$\sim$0.6 emission--line galaxies, which represents half that found in present--day spirals \citep{2008A&A...492..371R}. 

However, assuming a change in log(M$_{stellar}$/L$_K$) with redshift in the way described by \citet{2004ApJ...608..742D}, but no evolution in the K--band TFR as found in the present work, we obtain that the spiral intermediate--mass galaxies have doubled their stellar masses since z=1.25, which agrees rather well with expectations from stellar mass density studies \citep{2004ApJ...608..742D}. From the mass--to--light ratios in u, r, i and K--bands of \citet{2007A&A...465..417W} (they found a similar result for the K--band than that of \citet{2004ApJ...608..742D} in the last 9 Gyrs), but considering our growth in stellar mass by a factor 2, the evolution in the optical bands would be: $\Delta$M$_u$=$-$1.3; $\Delta$M$_r$=$-$0.79; $\Delta$M$_i$=$-$0.67 (for a IMF of Salpeter without absorption), which are consistent with the result found in the present work. Assuming at least the same evolution in the K--band magnitude at z=1.25 as that found by \citet{2008A&A...484..173P} at z=0.6, the stellar mass should have grown at least by a factor of 3.5 applying the evolution in log(M$_{stellar}$/L$_K$) found by \citet{2004ApJ...608..742D}. However, the stellar mass should grow by a factor of 2.6 since z=1.25, if the change in the gaseous O/H abundance were attributed to all gas transformed in stars \citep{2008A&A...492..371R}. Given the short lives of the most massive stars responsible for the enrichment of the ISM (interstellar medium), and the time required for cooling the gas to form new stars, the change in O/H in the last 8.6 Gyr will not be only due to a rise in O, but also to a drop in H. Then, likely the change in stellar mass derived from the metallicity evolution is an upper limit.

The stellar mass can also increase by galaxy merger processes or accretion of small galaxies. In the first case, the galaxy disc would be destroyed if the mass of the merger were larger than $\sim$15$\%$ of the parent galaxy mass. The accretion of small galaxies, although a probable contribution in increasing the stellar mass, it is not enough for explaining a growth by a factor 3.5. Hence, our result of a double stellar mass since z=1.25 seems more plausible and consistent. Likely the increase in stellar mass is due to gas transformed in stars, consistent with higher SFR density at z$\sim$1$-$2 than today \citep{2009ApJ...705..936B}. This gas could be present either in the disc, the halo or both places. In a recent work, \citet{2010Natur.463..781T} found empirical evidence that the larger star formation rates at z$\sim$1--2 are the consequence of large molecular gas reservoirs (both due to continuous gas inflow from the halos, and minor but not major mergers) and not of a larger star formation efficiency than that at z$\sim$0. Nevertheless, doubling the amount of gas at z=1.25, to account for doubling the stellar mass at z=0, would result in higher extinctions in the optical bands than those derived from the \citet{1998AJ....115.2264T} equations used in this work, because they have been obtained using local galaxy samples. Were this the case, the evolution found in the optical bands would be larger than the evolution observed in this work. 

\section{Conclusions}

We have studied the evolution of the TFR in B, V, R, I and K$_{\rm S}$--bands, using two sample of galaxies in the GSS, one in the redshift range 0.1$<$z$<$0.3 as representative of local galaxies; and the high redshift sample in the redshift range 1.1$<$z$<$1.4. The rotation velocity of all 241 spiral galaxies were measured from optical lines widths, using DEEP2 spectra. Morphology was obtained from HST images, and the absolute magnitudes were derived from a carefully study of the k--correction made with a large set of photometric information. The results of this study can be summarized as follows:

We analized four sets of templates and we concluded that the SED of a spiral galaxy is better reproduced by the nonnegative linear combination of five templates based on the \citet{2003MNRAS.344.1000B} stellar evolution synthesis codes obtained by {\tt kcorrect}. However, when the data include a noisy band, the k--corrections calculated by this code are not suitable. Also, we find that the most reliable k--correction is obtained from information in a observed band that roughly match the rest--frame band. In this case, the rest--frame magnitude in the noisy band is better calculated directly from the best--fit template. 

When the observed photometry does not match the rest--frame bands, it is possible to calculate the rest--frame magnitudes via an interpolation method, as long as enough information at larger and shorter wavelengths is available. However, an extrapolation method is not reliable for galaxies at z$\sim$1.25 since the results can vary as more than one magnitude mainly due to a worse fit of the template.

New local TFRs were constructed increasing the sample of our previous work \citep{2009fernandez} in the redshift range 0.1$<$z$<$0.3. We obtained a similar difference in the zero--point of the TFR for all bands when we compare with the \citet{2001ApJ...563..694V} local relations, that we attributte to the ratio between rotation velocity and line width.

We derived the high redshift 1.1$<$z$<$1.4 TFRs fixing the slope to the local one in each band, to study the evolution of the zero--points. We confirm the evolution in the B--band TFR found in our previous work \citep{2009fernandez}, in the sense that galaxies were brighter in the past, and no evolution in the K$_{\rm S}$--band TFR. Furthermore, for the other bands we find a gradual evolution, with a change in luminosity more noticiable as the band is bluer.

We study the slope of the local K$_{\rm S}$--band TFR comparing with Millenium simulated data. We obtained that the slope changes depending on the luminosity range used in the fit. Considering this effect, we calculated new local K$_{\rm S}$--band TFR from the Millenium data in the same luminosity range than our high redshift sample, confirming again no evolution in the K$_{\rm S}$--band TFR.

Assuming the change in the stellar mass--to--light ratio M/L$_K$ found by \citet{2007A&A...465..417W}, consistent with that of \citet{2004ApJ...608..742D}, but no evolution in the K$_{\rm S}$--band, we obtain that galaxies would have duplicated their stellar mass in the last 8.6 Gyr. Likely, this growth is mainly due to the gas been transformed into stars, as supported by the change in the gaseous O/H abundance found by \citet{2008A&A...492..371R}. From the mass--to--light ratios derived by \citet{2007A&A...465..417W} for the optical bands (IMF of Salpeter), and assuming this change in stellar mass, we infer an evolution in the optical luminosity similar to that found in the present work. However, the luminosity and colour evolution simulated by \citet{2007A&A...465..417W} is not consistent with that found here and would imply a growth of $\sim$6 in the stellar mass from z=1.25 to z=0. 

Therefore, the galaxies, despite having doubled their stellar mass in the last 8.6 Gyr, are nowadays fainter and redder in the optical bands, which can be attributed to an ageing of the stellar populations as consequence of the star formation density decrease observed in the last 8.6 Gyr. This is supported by the change found in the (V--K$_{\rm S}$) and (R--I) colours (for a fixed velocity), in the sense that galaxies were bluer in the past. 

 \begin{acknowledgements}

This work was supported by the Spanish {\em Plan Nacional de Astronom\'ia y Astrof\'isica} under grant AYA2008--06311--C02--01. We thank the DEEP2 group for making their catalogues and data publicly available. Funding for the DEEP2 survey has been provided by NSF grants AST95--09298, AST--0071048, AST--0071198, AST--0507428, and AST--0507483, as well as NASA LTSA grant NNG04GC89G. The work is based on observations obtained at the Canada--France--Hawaii Telescope (CFHT) which is operated by the National Research Council of Canada, the Institut National des Sciences de l'Univers of the Centre National de la Recherche Scientifique of France, and the University of Hawaii.

This study makes use of data from AEGIS, a multiwavelength sky survey conducted with the Chandra, GALEX, Hubble, Keck, CFHT, MMT, Subaru, Palomar, Spitzer, VLA, and other telescopes and supported in part by the NSF, NASA, and the STFC.

The Millenium Simulation database used in this paper and the web application providing online access to them were constructed as part of the activities of the German Astrophysical Virtual Observatory.


 \end{acknowledgements}

\begin{appendix}

\section{Deriving the k--corrections}

\begin{figure*}[h!t]
\centering
\includegraphics[scale=0.44]{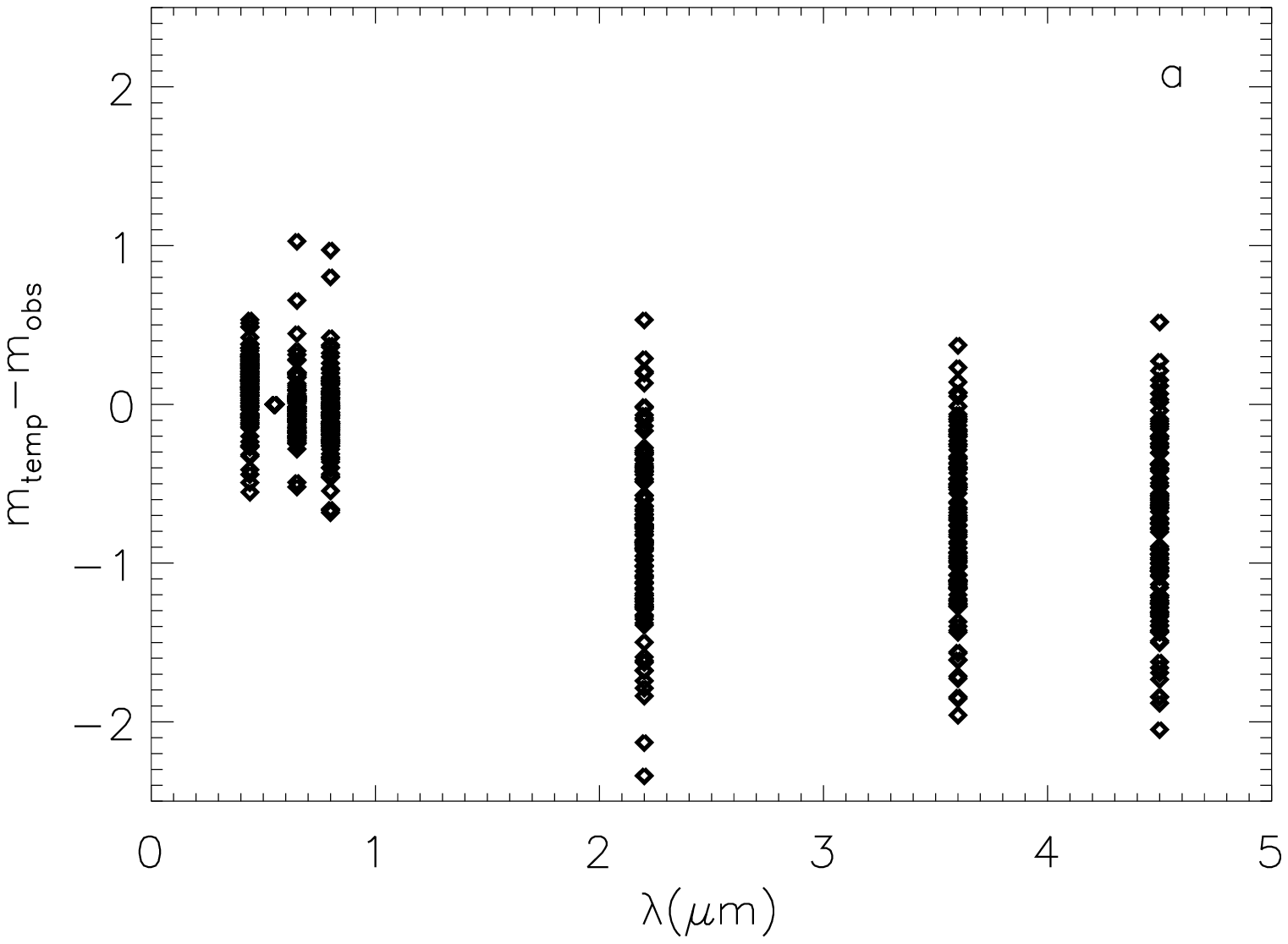}
\includegraphics[scale=0.44]{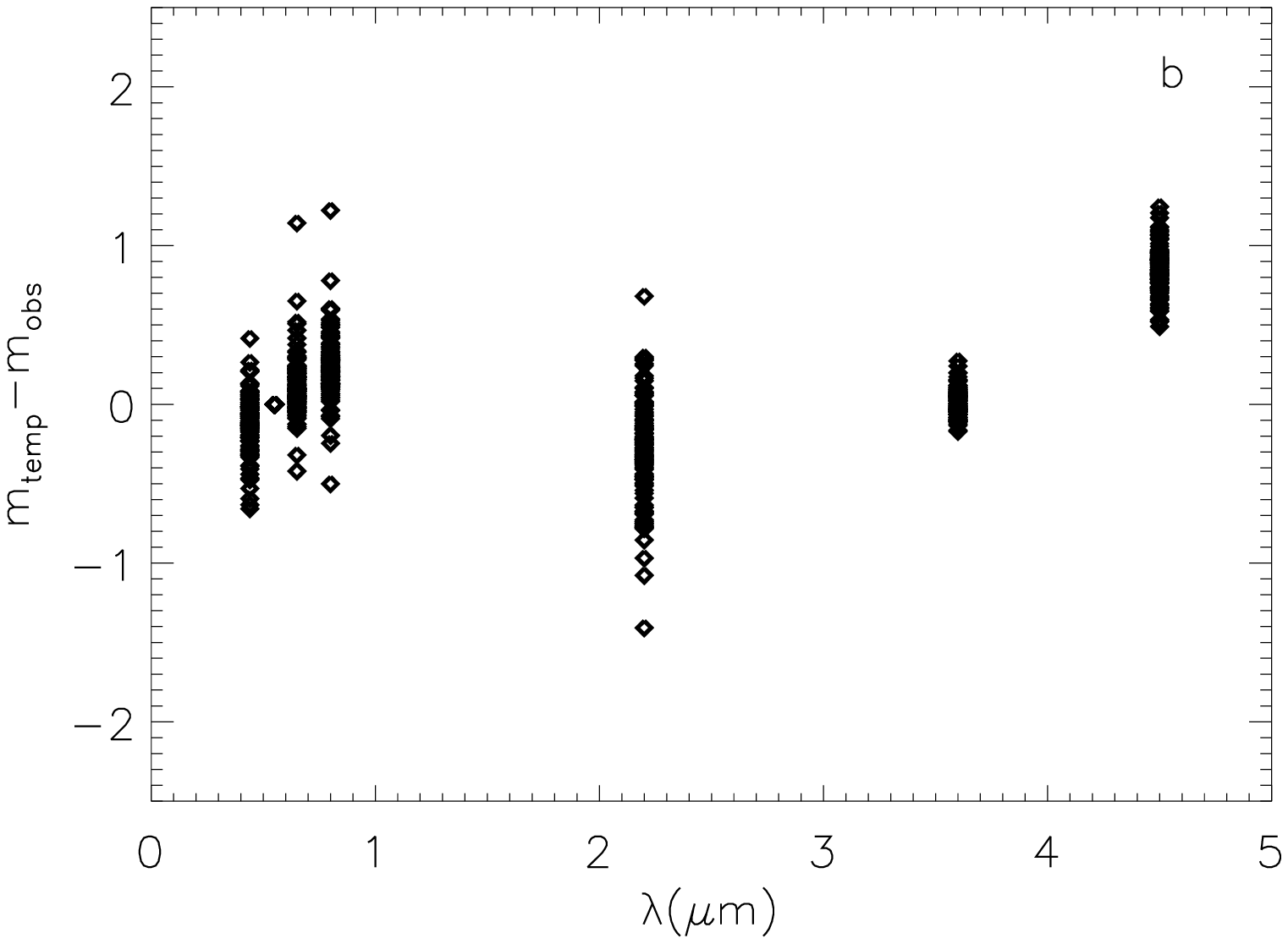}
\includegraphics[scale=0.44]{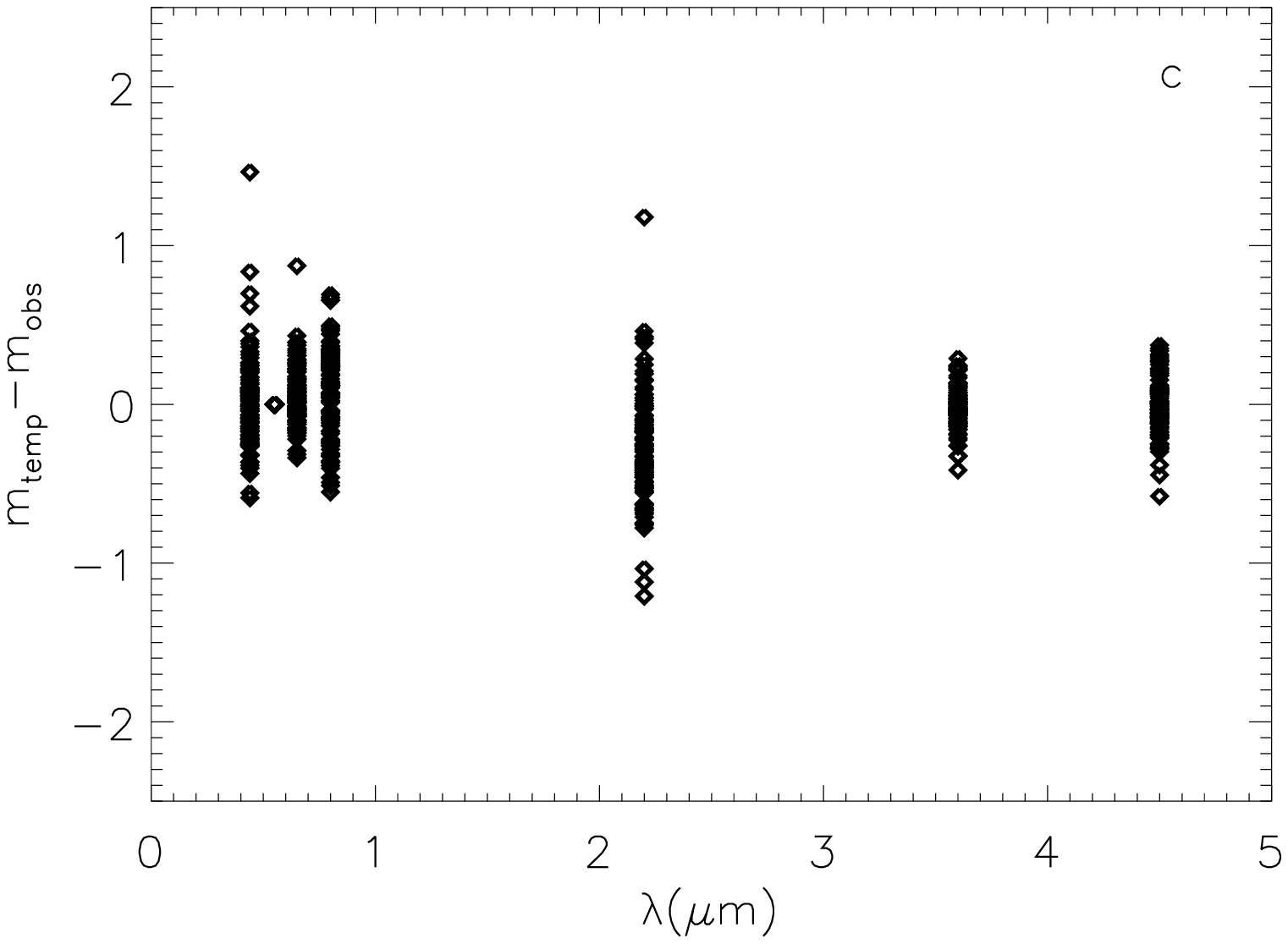}
\includegraphics[scale=0.44]{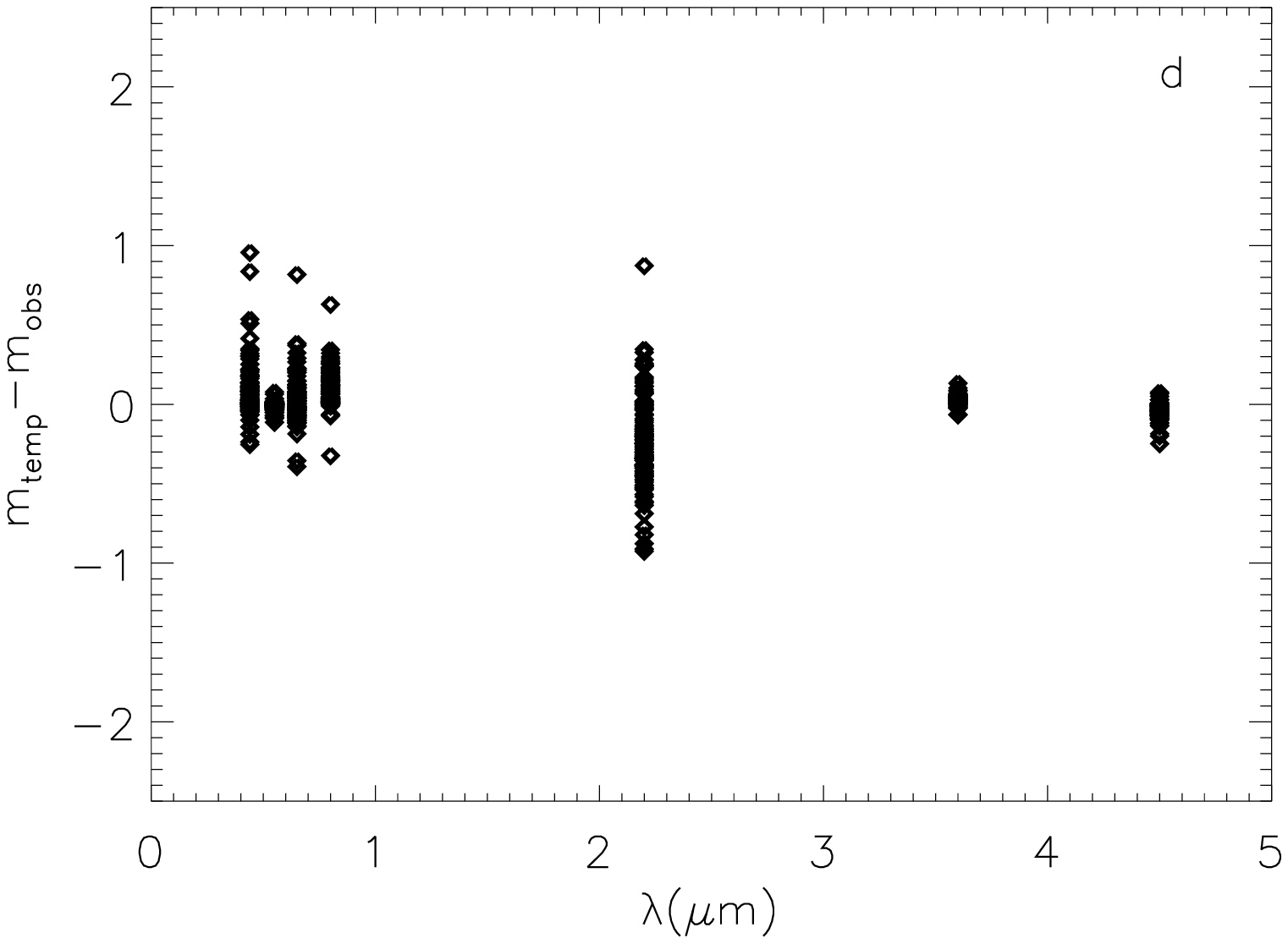}
\caption{Differences between the B, V, R, I, K$_{\rm S}$, IRAC1 and IRAC2--band magnitudes derived from the best--fit template normalized to the observed V--band magnitude (m$_{\rm temp}$), and the corresponding observed magnitudes (m$_{\rm obs}$), for our high--redshift sample. a) Templates of SWIRE. b) Templates of Poggianti. c) Templates of BC03. d) Templates provided by {\tt kcorrect}.}
\label{appfig}
\end{figure*}

In this section we present a systematic study using various sets of templates and methods to establish the most reliable procedure to do the k--correction on our data.

For this purpose, we have performed the following tests:

(i) We compare the results obtained fitting different sets of spiral templates to the data, to establish the set of templates that better fits our high redshift sample (see section A.1).

(ii) We estimate how good is the rest--frame photometry reproduced by the templates, since our photometry does not match the rest--frame optical magnitudes, and the K$_{\rm S}$--band errors are large. With this aim, we calculate the best--fit template of z=1.25 galaxies with a known SED (i.e.: templates of local galaxies redshifted to z=1.25), from the photometric information in the same bands and using the same weights that we have in our real sample. Then, we obtain the differences between the rest--frame optical magnitudes calculated from the original rest--frame SED, and those from the best--fit template (see section A.2).

(iii) We compare the accuracy of the k--corrections derived using this interpolation method with those obtained using an observed--band that roughly match the rest--frame magnitude (see section A.3).

\subsection{Best set of templates}

To evaluate the templates that better fit each of our galaxies, we used the code {\tt bpz}, which implements the bayesian photometric redshift method described in \citet{2000ApJ...536..571B}. We restrict the code redshift range to 0.95--1.4, where all the galaxies of our sample are located. This code was applied using four different sets of spiral galaxy templates:

(i) The templates of SWIRE, formed by 7 spirals ranging from early to late types (S0--Sdm), and generated with the GRASIL code \citep{1998ApJ...509..103S}.

(ii) The templates of Poggianti \citep{1997A&AS..122..399P}, that consist on models of Sa and Sc spirals computed for various redshifts from 0 to 3.

(iii) The \citet{2003MNRAS.344.1000B} (hereafter BC03) collection of galaxy templates used by \citet{2003AAS...202.5102T} in the analysis of SDSS (Sloan Digital Sky Survey) galaxy spectra. The library includes 39 templates, obtained from 13 models and three metallicities Z=0.008, 0.02, and 0.05 (see Table A.1 for the description of the 13 models).

(iv) The resulting templates obtained from the routine {\tt kcorrect} \citep{2007AJ....133..734B} which are a nonnegative linear combination of five templates based on the BC03 stellar evolution synthesis code.

For each galaxy, we calculate the redshifted best--fit template as follows:

\begin{equation}
  {\lambda} = {\lambda_{0}} \ (1+z) 
\end{equation}

\begin{equation}
  f_{\lambda} = \frac{f_{\lambda_{0}}}{(1+z)} .
\end{equation}

where ${\lambda_{0}}$ is the wavelength and $f_{\lambda_{0}}$ is the flux density in $\rm ergs \ s^{-1} \ cm ^{-2} \ \AA^{-1}$ of the template at z=0. Then, we normalize to the V--band to obtain the SED of each galaxy. As this template is the best representation of the photometry observed, we can know what set of templates reproduces better our data by calculating the difference between the observed magnitude $m_{obs}$ and the magnitude $m_{temp}$ obtained from the redshifted template, in each band. To calculate $m_{temp}$, we project the template into the filter $Q$, using the expresion:

\begin{equation}
  m_Q = -2.5 \ \log \ \left(\frac {\displaystyle\sum_l  (\lambda_{l+1} \ - \ \lambda_l) \ {\lambda_l} \ {f_{\lambda_l} \ T_{Q \lambda_{l}}}} {\displaystyle\sum_l (\lambda_{l+1} \ - \ \lambda_l) \ \frac{c} {{\lambda_l}}  \ f_{AB} \ (\nu) \ T_{Q \lambda_{l} }}\right) . 
\end{equation}

where $T_{Q \lambda_{l} }$ is the filter transmission in the Q--band at the wavelength $\lambda_{l}$, $f_{AB} \ (\nu)$ = 3.631 x 10$^{-20}$ ergs s$^{-1}$ cm$^{-2}$ Hz$^{-1}$ is the flux density of the AB standard source, and $c$ is the speed of light.

\begin{table}
\caption{Description of the templates from BC03 and number of galaxies (N$_{\rm G}$) in our sample that were fitted to each one by {\tt bpz}. Each model was derived for 3 metallicities: Z = 0.008, 0.02, 0.05.}             
\label{table:1}      
\centering                          
\begin{tabular}{c c c c}
\hline
\\
{\bf Templates} & {\bf Model description} & {Age} & {\bf N$_{\rm G}$} \\\hline
1--3 & Constant star formation & 6 Gyr & 36 \\
4--6 & Instantaneus--burst & 5 Myr & 0 \\
7--9 & Instantaneus--burst & 25 Myr & 1 \\
10--12 & Instantaneus--burst & 100 Myr & 21 \\
13--15 & Instantaneus--burst & 290 Myr & 32 \\
16--18 & Instantaneus--burst & 640 Myr & 7 \\
19--21 & Instantaneus--burst & 900 Myr & 5 \\
22--24 & Instantaneus--burst & 1.4 Gyr & 3 \\
25--27 & Instantaneus--burst & 2.5 Gyr & 0 \\
28--30 & Instantaneus--burst & 5 Gyr & 1 \\
31--33 & Instantaneus--burst & 11 Gyr & 0 \\
34--36 & Exp. declining SFR ${\tau}_{SFR}$ = 5 Gyr & 12 Gyr & 8 \\
37--39 & Exp. declining SFR ${\tau}_{SFR}$ = 9 Gyr & 12 Gyr & 28 \\
\hline
\end{tabular}
\end{table}

\begin{figure}[!t]
   \centering
   \includegraphics[angle=0,width=7.2cm]{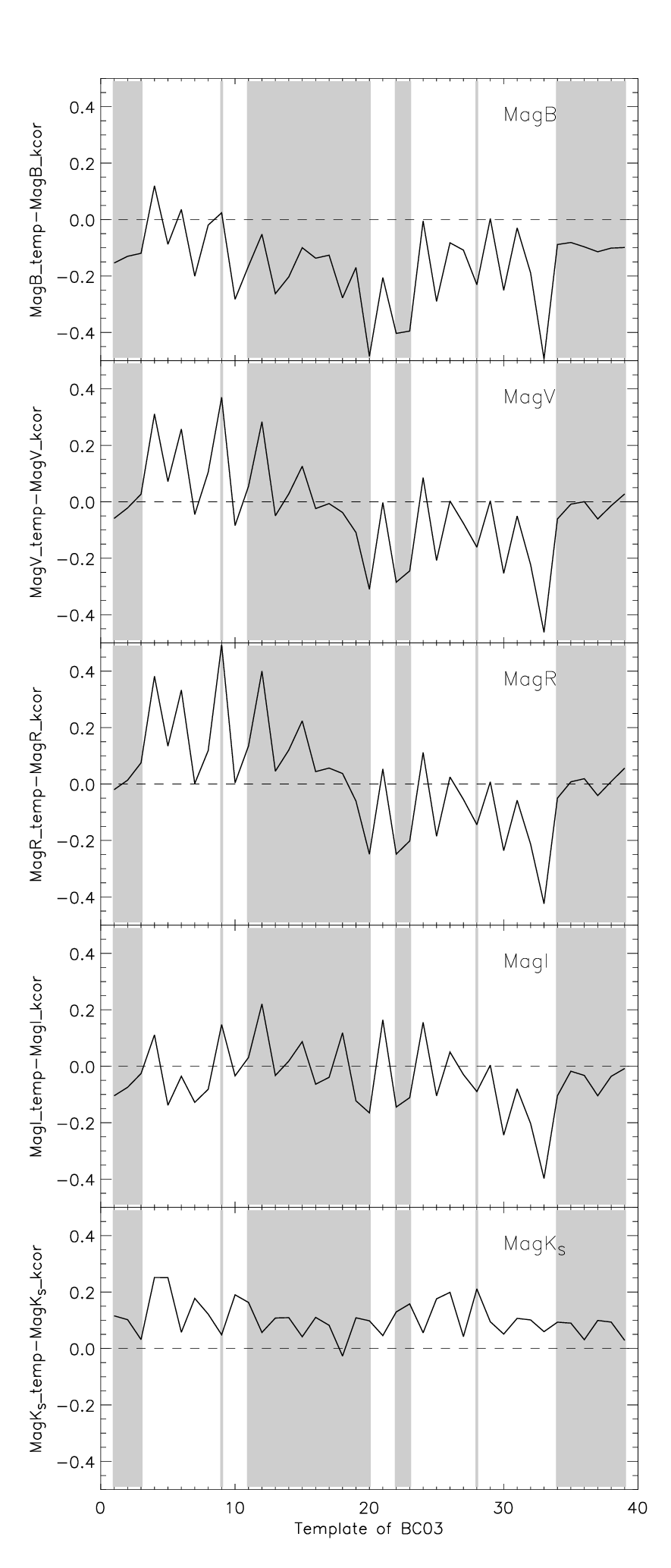}
      \caption{Difference between the magnitudes calculated from each rest--frame template of BC03 and from the best--fit template (at z=0) calculated by {\tt kcorrect} from the photometry of the redshifted template at z=1.25. See Table A.1 for the description of the templates from BC03. The grey zones represent the templates that were fitted by {\tt bpz} for some of our galaxies.}
   \end{figure}

In Fig. A.1 we have represented the result for all galaxies in each set of templates. For our objects, with a median redshift of $<$z$>$=1.25, B, V, R and I--bands correspond to UV rest--frame wavelenghts, while K$_{\rm S}$, IRAC1 and IRAC2--bands correspond to NIR rest--frame wavelenghts. For all sets of templates, the larger difference is found in the K$_{\rm S}$--band due to the photometric errors in this magnitude (the fits of {\tt bpz} and {\tt kcorrect} are weighed by the photometric errors). However, there is a clear difference between UV and NIR resulting from the comparison with the SWIRE templates (Fig. A.1a). For this set, the galaxies are bluer than the templates. On the contrary, for the Poggianti's templates (Fig. A.1b), the opposite happens, i.e. the galaxies are redder than the templates. However, with the set of BC03 (Fig. A.1c), the agreement between observed--frame bands and the magnitudes obtained from the templates, are better. The combination of the templates of BC03 obtained through {\tt kcorrect} (Fig. A.1d) provides a similar result, but with smaller dispersions. In this case, the template supplied by the code does not need to be normalized to V--band. In addition to the best--fit template, {\tt kcorrect} gives the k--correction for each band used in the fit. This k--correction is the difference between the redshifted and rest--frame band, both calculated from the best--fit template. Then our rest--frame magnitudes would be obtained by subtracting this k--corrections to the observed magnitudes. However, the photometric errors of the observed quantities propagate to the rest--frame magnitudes obtained in this way. This effect spoils colour evolution studies if one of the observed bands have substantially larger errors than the others. This is the case of our K$_{\rm S}$--band data, so that we used another approach for deriving the rest--frame K$_{\rm S}$ magnitudes. Since the observed IRAC2--band roughly match the rest--frame K$_{\rm S}$--band, and the fit of the template is quite good in the wavelength range corresponding to IRAC2 (see Fig. A.1d), we used directly the magnitudes obtained in this band from the rest--frame best--fit template instead of k--correcting the data. For the optical bands, the rest--frame magnitudes calculated from the best--fit template or using the k--corrections of {\tt kcorrect} are equivalent in terms of the TFR, because the average difference between both methods is almost null (see Fig. A.1d). Then, as in K$_{\rm S}$--band, the best--fit templates obtained through {\tt kcorrect} will be used, but not the k--corrections provided by this software.

\subsection{The reliability of the photometry}

\begin{figure}[!t]
   \centering
   \includegraphics[angle=0,width=8.2cm]{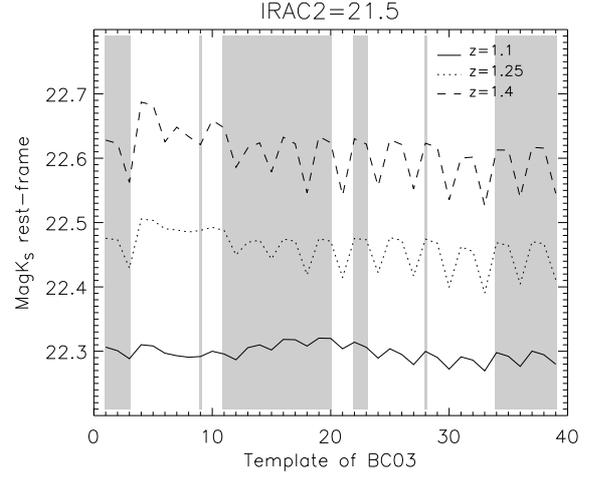}
      \caption{Rest--frame K$_{\rm S}$--band magnitude calculated from each of the templates from BC03 normalized to an IRAC2 magnitude of 21.5 in the redshifts z=1.1, 1.25 and 1.4. The description of the templates is the same as in Fig. A.2.}
   \end{figure}

Since our photometry does not match the rest--frame optical magnitudes, we need to know how reliably {\tt kcorrect} can reproduce the optical SED of a galaxy at 1.1$<$z$<$1.4 from the photometric information available. With this aim, we have redshifted the templates from BC03 to our median redshift, z=1.25, and we have calculated the magnitude in B, V, R, I, K$_{\rm S}$, IRAC1 and IRAC2--bands. Then, we have introduced this photometry in {\tt kcorrect} to obtain the best--fit template. We have adopted the average magnitude error of our data as the standard deviation used by the code in every band. Thus, the bands are weighted as the real data. Then, we calculated the difference in magnitude between the non--redshifted templates from BC03 and the resulted template of {\tt kcorrect} on the rest--frame B, V, R, I and K$_{\rm S}$--bands. Such differences are represented in Fig. A.2 and provide a first order estimation of the accuracy of the rest--frame magnitudes obtained in the previous section. As we have not introduced photometric information that match the optical rest--frame wavelength range, the {\tt kcorrect} template is providing interpolated magnitudes for these bands. The general tendency for the B--band is that {\tt kcorrect} provide a fainter rest--frame magnitude that the template from BC03 does, while the K$_{\rm S}$--band rest--frame magnitudes are brighter. However, the average difference for V, R, and I--bands is $\sim$0 when considering the full set of templates. For the constant star formation model (first 3 templates) and for the two exponential models (last 6 templates), the error in the SED reconstructed by {\tt kcorrect} is $\sim$0.1, whereas the worst fit is obtained with a simple stellar population. In general, the fit of more recent bursts provides optical rest--frame magnitudes brighter than the real magnitudes of the templates from BC03, while the optical information of the oldest burst reconstructed by {\tt kcorrect} tends to be fainter. For the simple stellar populations that fit some of our galaxies, the errors in some cases are $\sim$0.4. A recalculation of the k--corrections in our previous work \citep{2009fernandez}, adding the K$_{\rm S}$--band photometry to the B, R and I--bands used in this work, shows that our previous rest--frame magnitudes were overestimated, especially in the I--band. Moreover, the best--fit template of BC03 provided by {\tt bpz} is different if we use only the information in the B, R and I--bands, or if we introduce also the infrared information. We repeated the procedure used in the Fig. A.2, but introducing only the optical bands (B, V, R and I) in the {\tt kcorrect} fit, so that the rest--frame magnitudes are extrapolated rather than interpolated. The error in the k--corrected magnitudes is now between 0.5 and 1.5 mag for most of the templates. Then, the extrapolation method is not suitable for high redshift galaxies.

\subsection{Comparison of methods}

\begin{figure}[!t]
\centering
\includegraphics[scale=0.44]{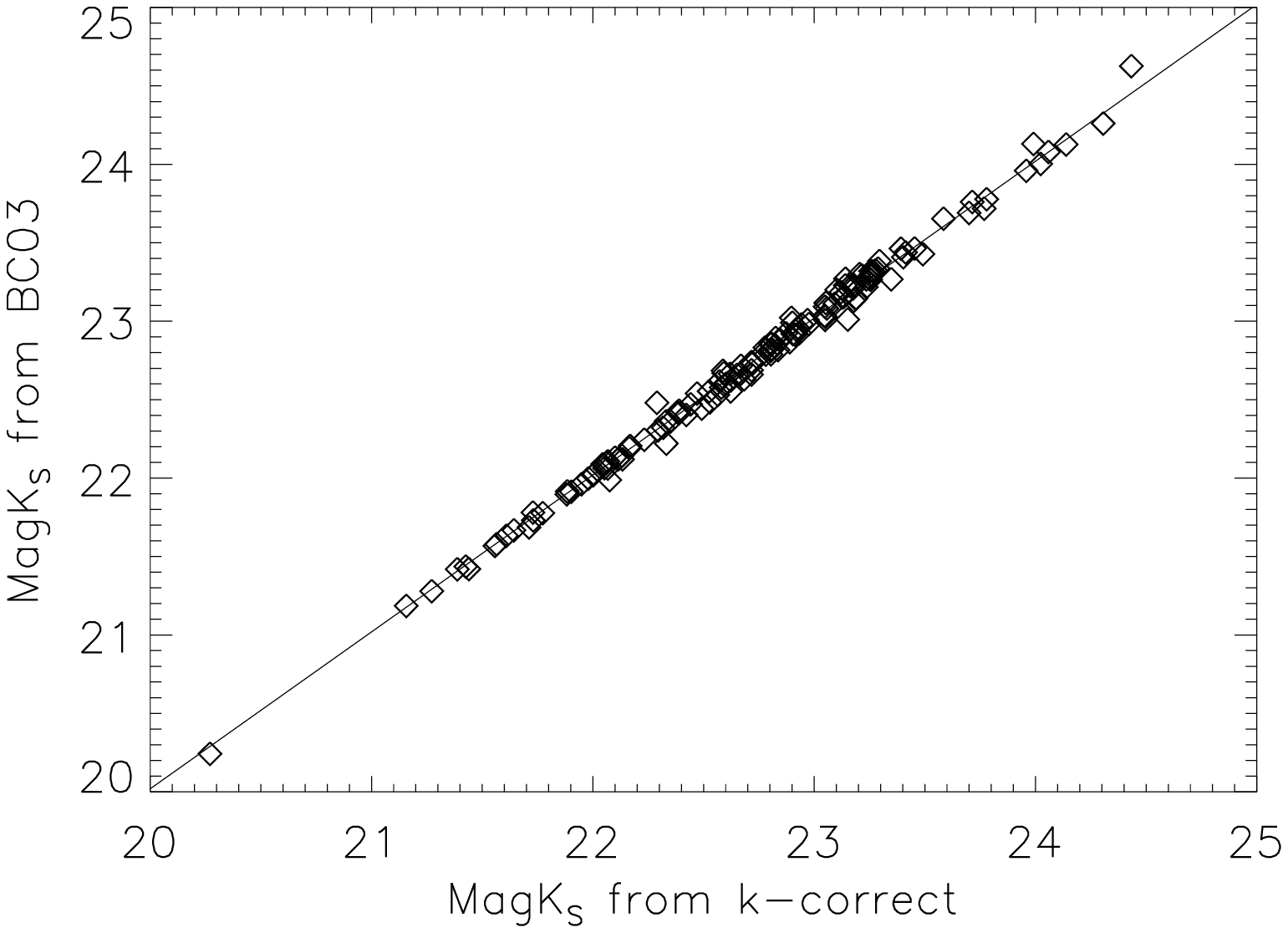}
\includegraphics[scale=0.44]{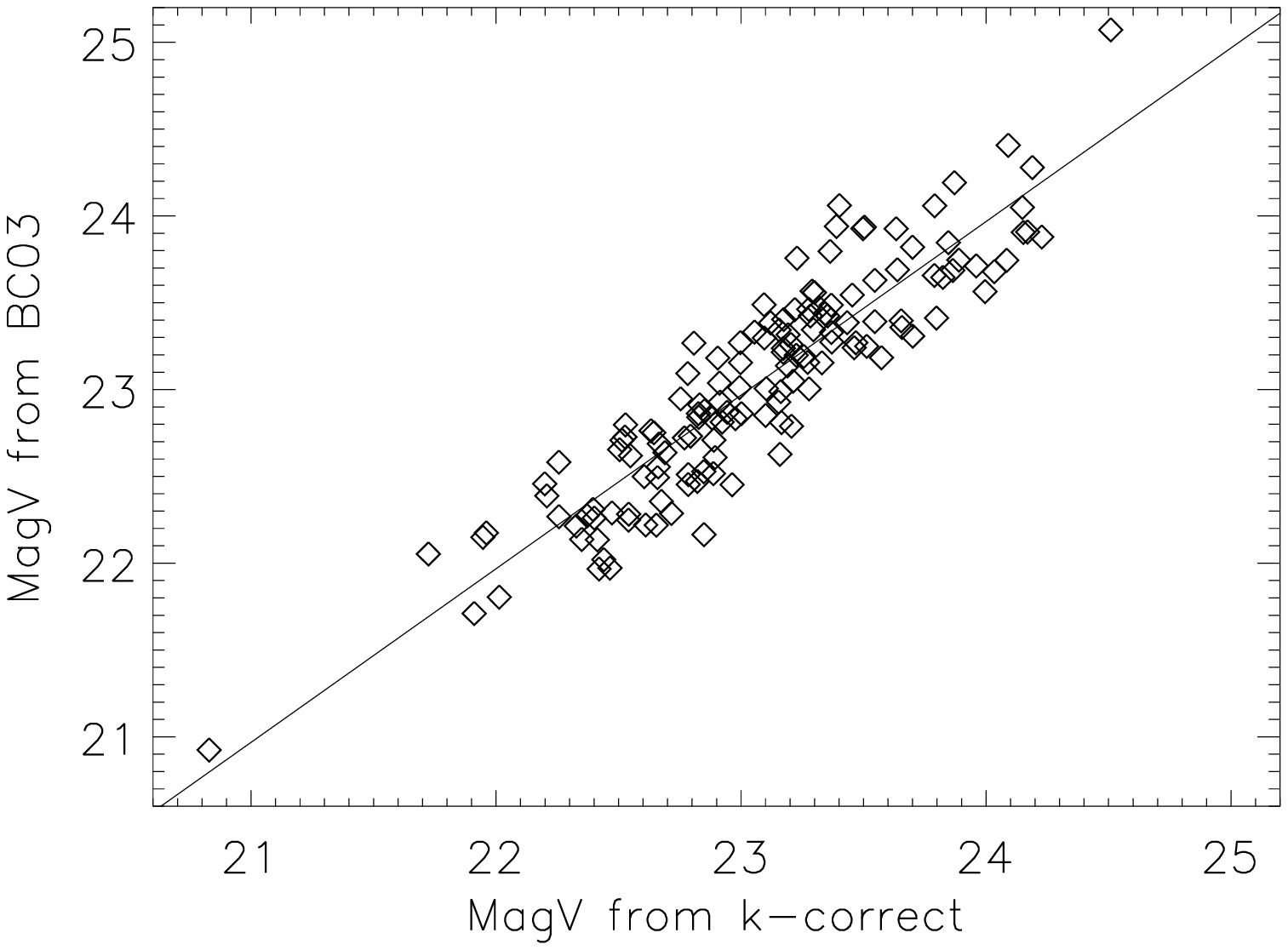}
\caption{Comparison between the k--corrected magnitude obtained from {\tt kcorrect} best--fit template and from the best--fit template of BC03. For K$_{\rm S}$--band (up), the normalization of the template from BC03 was done to the observed IRAC2--band. For V--band (down), the template of BC03 was normalized to the observed V--band. The solid line shows the least--square fits with slope 1.}
\end{figure}

\begin{figure*}[!t]
\centering
\includegraphics[scale=0.44]{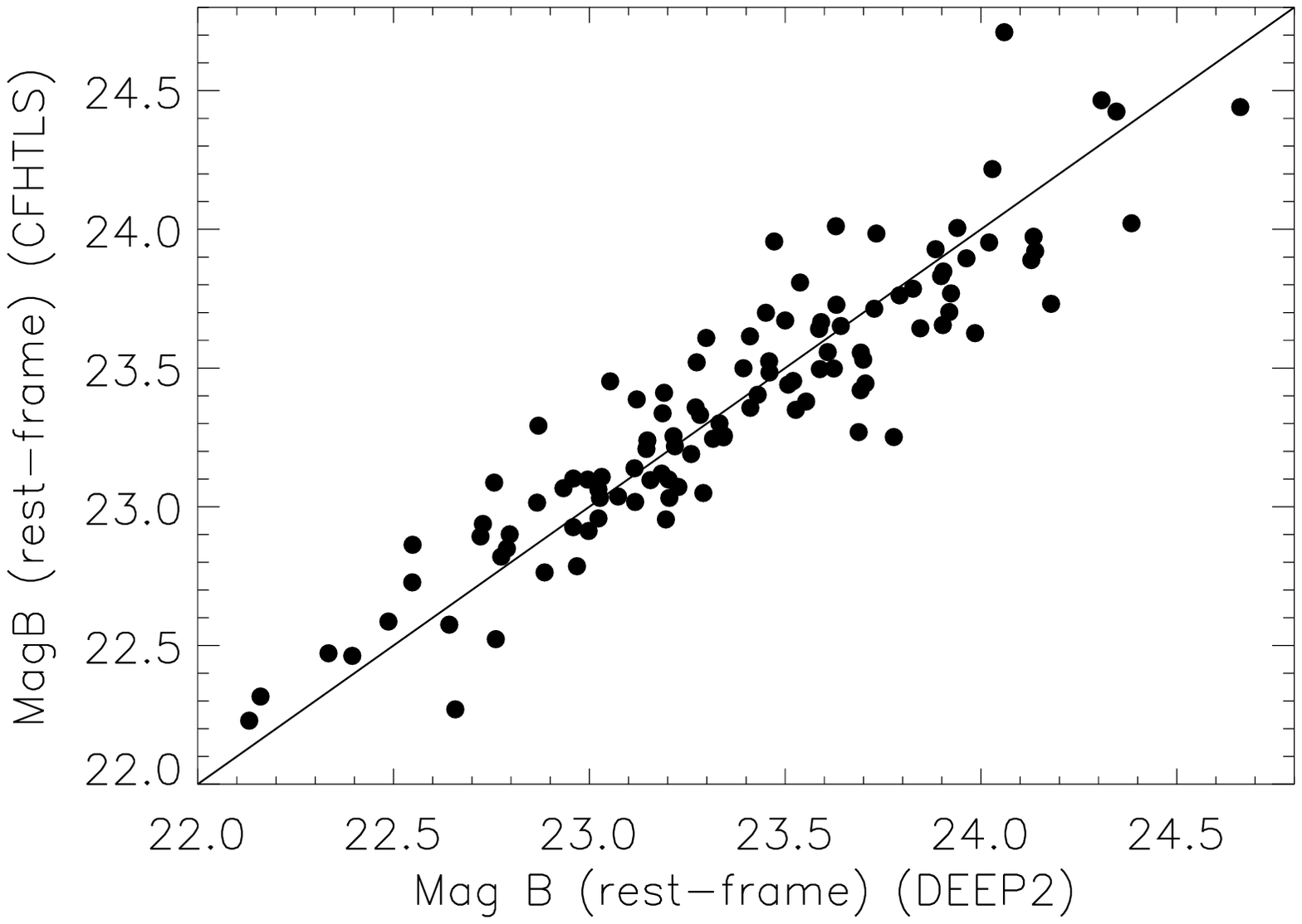}
\includegraphics[scale=0.44]{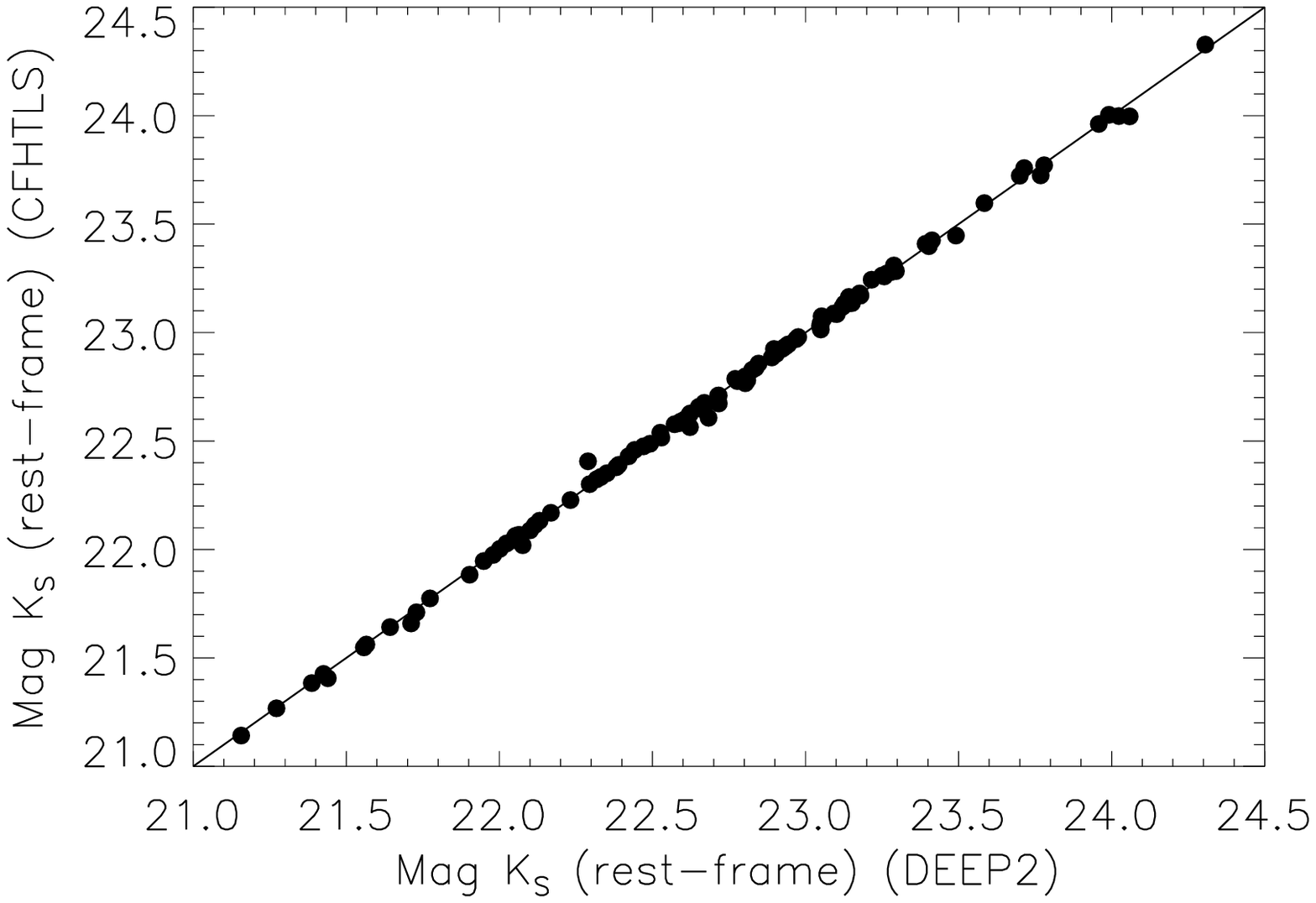}
\caption{Comparison between the rest--frame B--band (left) and K$_{\rm S}$--band (right) magnitudes calculated from the best--fit template obtained using the g, r, i, z, K$_{\rm S}$, IRAC1 and IRAC2--bands aperture magnitudes in 1.5" aperture, and those obtained using the B, V, R, I, K$_{\rm S}$, IRAC1 and IRAC2 total magnitudes. The solid lines are the one to one relations.}
\end{figure*}

Now, we are going to study the k--corrections obtained using an observed band that roughly match the rest--frame magnitude. In our redshift range, IRAC2 gives information about the SED region corresponding to rest--frame K$_{\rm S}$--band. Hence, if we normalize the redshifted best--fitting template to the IRAC2--band, and we calculate the K$_{\rm S}$ magnitude from this template at z=0, we could have more accuracy in the rest--frame K$_{\rm S}$--band magnitudes. The main difficulty using this procedure is the error in the choice of the template. In order to quantify this error, we are going to compare the k--corrected K$_{\rm S}$--band magnitude obtained from all templates of BC03 redshifted to z=1.1, 1.25 and 1.4, and normalized to the same IRAC2 magnitude. In Fig. A.3 we have represented the results. The larger difference between two templates is ${\sim}$0.1 magnitudes at z=1.4. We did the same procedure with the SWIRE templates, obtaining a similar result. Then, using photometry that roughly match the rest--frame K$_{\rm S}$--band, the k--correction is not too dependent of the template, mainly due to the smaller difference in the infrared SEDs between early and late--types spirals \citep[see, for example,][]{2001MNRAS.326..745M}. However, for optical B, V, R, and I--bands, this does not hold. Some of the galaxies have J--band photometry from Palomar data, that match the rest--frame V--band magnitude at z=1.25. Following the same procedure that the used with K$_{\rm S}$--band (Fig. A.3), but normalizing the templates to m$_{\rm J}$=22.5, we obtained that the maximum difference between the rest--frame V--band of two different templates was ${\sim}$0.3. Since the errors in the J--band are larger than the IRAC2 errors, this method is not as good for the V--band as for the K$_{\rm S}$--band k--correction.

Finally, we apply both methods to our sample of galaxies instead of redshifted templates. In Fig. A.4 we have represented the rest--frame magnitudes calculated from the best--fit template of BC03 normalized to the IRAC2--band, versus the rest--frame magnitudes calculated from the best--fit template of {\tt kcorrect}. In addition, we have represented the result in the V--band, but the template from BC03 was normalized to the observed--frame V--band instead the J--band. For the K$_{\rm S}$--band, the difference between both methods is negligible, but we found a larger scatter between the values of rest--frame V--band. Even so, both values are going to provide us the same mean result because the averaged differences are almost null. 

\subsection{Aperture effects}

All the photometry used in this work is part of the AEGIS project and has already been corrected for aperture and PSF effects. However, the magnitudes derived at the same aperture for all bands are not available. Then, we have chosen to compare the results obtained in this work with those obtained using other optical photometry. With this aim we have used the aperture photometry in the g, r, i and z--bands, which belong to the CFHTLS data and are part of AEGIS too. The aperture used by this team is 1.5", enough to cover the whole galaxies in the high redshift sample. Then, we have calculated the best--fit template using the aperture photometry in the g, r, i, z, K$_{\rm S}$, IRAC1 and IRAC2 bands, with an aperture of 1.5" (for the noisy K$_{\rm S}$--band we used mag$\_$AUTO). In Fig A.5 we have represented the rest--frame B and K$_{\rm S}$--bands calculated this way, and using the total magnitudes in the B, V, R, I, K$_{\rm S}$, IRAC1 and IRAC2--bands, which were determined as described in the section 3.1 of the paper. The dispersion obtained in the rest--frame B--band is mainly due to the interpolation, the same result that we obtained previously (Fig.A.2). For the rest--frame K$_{\rm S}$--band, we obtain very good agreement between the magnitudes derived from aperture and total photometry, since these k--corrections are basically dependent to the IRAC2--band, that roughly match the rest--frame K$_{\rm S}$--band. In this case, the k--correction is not too dependent of the template, as we explained in the section A.3 (and Fig.A.3).

\end{appendix}


\begin{thebibliography}{}

 \bibitem[Bamford et al.(2006)]{2006MNRAS.366..308B} Bamford, S.~P., Arag{\'o}n-Salamanca, A., \& Milvang-Jensen, B.\ 2006, \mnras, 366, 308
\bibitem[Barmby et al.(2008)]{2008ApJS..177..431B} Barmby, P., Huang, J.-S., Ashby, M.~L.~N., et al.\ 2008, \apjs, 177, 431
 \bibitem[Ben{\'{\i}}tez(2000)]{2000ApJ...536..571B} Ben{\'{\i}}tez, N.\ 
2000, \apj, 536, 571 
 \bibitem[Bertin \& Arnouts(1996)]{1996A&AS..117..393B} Bertin, E., \& Arnouts, S.\ 1996, \aaps, 117, 393
 \bibitem[Blanton \& Roweis(2007)]{2007AJ....133..734B} Blanton, M.~R., \& Roweis, S.\ 2007, \aj, 133, 734
 \bibitem[B{\"o}hm et al.(2004)]{2004A&A...420...97B} B{\"o}hm, A., Ziegler, B.~L., Saglia, R.~P., et al.\ 2004, \aap, 420, 97
\bibitem[Bosma(1981)]{1981AJ.....86.1825B} Bosma, A.\ 1981, \aj, 86, 1825 
\bibitem[Bouwens et al.(2009)]{2009ApJ...705..936B} Bouwens, R.~J., Illingworth, G. D., Franx, M., et al.\ 2009, \apj, 705, 936 
\bibitem[Bruzual \& Charlot(2003)]{2003MNRAS.344.1000B} Bruzual, G., \& Charlot, S.\ 2003, \mnras, 344, 1000 (BC03)
\bibitem[Bundy et al.(2006)]{2006bundy} Bundy, K., Ellis, R.~S., Conselice, C.~J., et al.\ 2006, \apj, 651, 120
\bibitem[Calzetti et al.(2000)]{2000ApJ...533..682C} Calzetti, D., Armus, L., Bohlin, et al. \ 2000, \apj, 533, 682 
 \bibitem[Coil et al.(2004)]{2004ApJ...617..765C} Coil, A.~L., Newman, J.~A., Kaiser, N., et al.\ 2004, \apj, 617, 765
 \bibitem[Conselice et al.(2005)]{2005ApJ...628..160C} Conselice, C.~J., Bundy, K., Ellis, R.~S., et al.\ 2005, \apj, 628, 160
\bibitem[Croton et al.(2006)]{2006MNRAS.365...11C} Croton, D.~J., Springel, V., White, S., et al.\ 2006, \mnras, 365, 11 

 \bibitem[Cuillandre et al.(2001)]{2001ASPC..232..398C} Cuillandre, J.-C., Starr, B., Isani, S., McDonald, J.~S., \& Luppino, G.\ 2001, ASPC, 232, 398
 \bibitem[Davis et al.(2003)]{2003SPIE.4834..161D} Davis, M., Faber, S.~M., Newman, J., et al.\ 2003, SPIE, 4834, 161
 \bibitem[Davis et al.(2007)]{2007ApJ...660L...1D} Davis, M., Guhathakurta, P., Konidaris, N., et al.\ 2007, \apjl, 660, L1
\bibitem[De Lucia \& Blaizot(2007)]{2007MNRAS.375....2D} De Lucia, G., \& Blaizot, J.\ 2007, \mnras, 375, 2
\bibitem[Drory et al.(2004)]{2004ApJ...608..742D} Drory, N., Bender, R., Feulner, G., et al.\ 2004, \apj, 608, 742 
\bibitem[Faber et al.(2003)]{2003SPIE.4841.1657F}  Faber, S.~M., Phillips, A.~C., Kibrick, R.~I., et al.\ 2003, SPIE, 4841, 1657
\bibitem[Fern\'andez Lorenzo et al. (2009)]{2009fernandez} Fern\'andez Lorenzo, M., Cepa, J., Bongiovanni, A., et al.\ 2009, \aap, 496, 389 
 \bibitem[Flores et al.(2006)]{2006A&A...455..107F} Flores, H., Hammer, F., Puech, M., Amram, P., \& Balkowski, C.\ 2006, \aap, 455, 107
\bibitem[Giovanelli et al.(1997)]{1997AJ....113...22G} Giovanelli, R., Haynes, M.~P., Herter, T., et al.\ 1997, \aj, 113, 22
\bibitem[Grocholski \& Sarajedini(2002)]{2002AJ....123.1603G} Grocholski, A.~J., \& Sarajedini, A.\ 2002, \aj, 123, 1603
\bibitem[Hammer et al.(2007)]{2007hammer}  Hammer, F., Puech, M., Chemin, L., Flores, H., \& Lehnert, M.~D.\ 2007, \apj, 662, 322 
\bibitem[Hewett et al.(2006)]{2006hewett} Hewett P. C., Warren S. J., Leggett S. K., \& Hodgkin S. T.\ 2006, \mnras, 367, 454
 \bibitem[Horne(1986)]{1986PASP...98..609H} Horne, K.\ 1986, \pasp, 98, 609
 \bibitem[Howarth et al.(1996)]{1996sun50} Howarth I.D., Murray J., Mills D., \& Berry D.S.\ 1996, Starlink User Note 50.21
 \bibitem[Kassin et al.(2007)]{2007ApJ...660L..35K} Kassin, S.~A., Weiner, B.~J., Faber, S.~M., et al.\ 2007, \apjl, 660, L35
\bibitem[Laurikainen et al.(2007)]{2006laurikainen} Laurikainen E., Salo H., Buta R., et al.\ 2007, IAU Symposium, 235, 36 
\bibitem[Mannucci et al.(2001)]{2001MNRAS.326..745M} Mannucci, F., Basile, F., Poggianti, et al.\ 2001, \mnras, 326, 745 
\bibitem[Masters et al.(2008)]{2008AJ....135.1738M} Masters, K.~L., Springob, C.~M., \& Huchra, J.~P.\ 2008, \aj, 135, 1738
 \bibitem[Mathewson et al.(1992)]{1992ApJS...81..413M} Mathewson, D.~S., Ford, V.~L., \& Buchhorn, M.\ 1992, \apjs, 81, 413
 \bibitem[Nakamura et al.(2006)]{2006MNRAS.366..144N} Nakamura, O., Arag\'on--Salamanca, A., Milvang--Jensen, B., et al.\ 2006, \mnras, 366, 144
 \bibitem[Navarro \& Steinmetz(2000)]{2000ApJ...538..477N} Navarro, J.~F., \& Steinmetz, M.\ 2000, \apj, 538, 477
 \bibitem[Poggianti(1997)]{1997A&AS..122..399P} Poggianti, B.~M.\ 1997, \aaps, 122, 399 

 \bibitem[Puech et al.(2008)]{2008A&A...484..173P} Puech, M., Flores, H., Hammer, F., et al.\ 2008, \aap, 484, 173
 \bibitem[Rix et al.(1997)]{1997MNRAS.285..779R} Rix, H.-W., Guhathakurta, P., Colless, M., \& Ing, K.\ 1997, \mnras, 285, 779
\bibitem[Rodrigues et al.(2008)]{2008A&A...492..371R} Rodrigues, M., Hammer, F., Flores, H. et al.\ 2008, \aap, 492, 371
 \bibitem[Schlegel et al.(1998)]{1998ApJ...500..525S} Schlegel, D.~J., Finkbeiner, D.~P., \& Davis, M.\ 1998, \apj, 500, 525
 \bibitem[Schechter(1980)]{1980AJ.....85..801S} Schechter, P.~L.\ 1980, \aj, 85, 801
\bibitem[Silva et al.(1998)]{1998ApJ...509..103S} Silva, L., Granato, G.~L., Bressan, A., \& Danese, L.\ 1998, \apj, 509, 103 
\bibitem[Simard(1998)]{1998ASPC..145..108S} Simard, L.\ 1998, Astronomical 
Data Analysis Software and Systems VII, 145, 108 
 \bibitem[Simard \& Pritcher (1998)]{1998simard} Simard, L., \& Pritcher, C.\ 1998, \apj, 505, 96
\bibitem[Springel et al.(2005)]{2005Natur.435..629S} Springel, V., et al.\ 2005, \nat, 435, 629
\bibitem[Tacconi et al.(2010)]{2010Natur.463..781T} Tacconi, L.~J., Genzel, R., Neri, R., et al.\ 
2010, \nat, 463, 781
 \bibitem[Tremonti et al.(2003)]{2003AAS...202.5102T} Tremonti, C.~A., Heckman, T.~M., Kauffmann, G., et al.\ 2003, AAS, 35, 770
 \bibitem[Tully \& Fisher(1977)]{1977A&A....54..661T} Tully, R.~B., \& Fisher, J.~R.\ 1977, \aap, 54, 661
 \bibitem[Tully et al.(1998)]{1998AJ....115.2264T} Tully, R.~B., Pierce, M.~J., Huang, J.-S., et al.\ 1998, \aj, 115, 2264
 \bibitem[Tully \& Pierce(2000)]{2000ApJ..533..744} Tully, R.~B., \& Pierce, M.~J.,\ 2000, \apj, 533, 744
\bibitem[van Starkenburg et al.(2006)]{2006A&A...450...25V} van Starkenburg, L., van der Werf, P.~P., Yan, L., \& Moorwood, A.~F.~M.\ 2006, \aap, 450, 25 
 \bibitem[Verheijen(2001)]{2001ApJ...563..694V} Verheijen, M.~A.~W.\ 2001, \apj, 563, 694
\bibitem[Vogt et al. (1996)]{1996vogt} Vogt, N.~P., Forbes, D.~A., Phillips, A.~C., et al.\ 1996, \apj, 465, 15
\bibitem[Vogt et al. (1997)]{1997vogt} Vogt, N.~P., Phillips, A.~C., Faber, S.~M., et al.,\ 1997, \apj, 479, 121
\bibitem[Weiner et al. (2006a)]{2006weinerI} Weiner, B.~J., Willmer, C.~N.~A., Faber, S.~M., et al.,\ 2006, \apj, 653, 1027
\bibitem[Weiner et al. (2006b)]{2006weinerII} Weiner, B.~J., Willmer, C.~N.~A., Faber, S.~M., et al.,\ 2006, \apj, 653, 1049
 \bibitem[Westera et al.(2002)]{2002A&A...389..761W} Westera, P., Samland, M., Buser, R., \& Gerhard, O.~E.\ 2002, \aap, 389, 761
\bibitem[Westera et al.(2007)]{2007A&A...465..417W} Westera, P., Samland, M., Kautsch, S.~J., Buser, R., \& Ammon, K.\ 2007, \aap, 465, 417
 \bibitem[Willick(1994)]{1994ApJS...92....1W} Willick, J.~A.\ 1994, \apjs, 92, 1
\bibitem[Ziegler et al.(2002)]{2002ApJ...564L..69Z} Ziegler, B.~L., B{\"o}hm, A., Fricke, K.~J., et al.\ 2002, \apjl, 564, L69
\bibitem[Zwaan et al.(1995)]{1995MNRAS.273L..35Z} Zwaan, M.~A., van der Hulst, J.~M., de Blok, W.~J.~G., \& McGaugh, S.~S.\ 1995, \mnras, 273, L35

\end{thebibliography}
\end{document}